\documentclass[aps,twocolumn,superscriptaddress,amsmath,amssymb,amsfonts,citeautoscript]{revtex4-1}
\usepackage{graphicx}
\usepackage{bm}
\usepackage{color}
\usepackage{epstopdf}
\usepackage{amsmath}
\usepackage{amssymb}
\usepackage{ulem}
\usepackage[urlcolor=blue,colorlinks=true,citecolor=blue,linkcolor=blue,pdfstartview={FitH},bookmarks=false]{hyperref}
\usepackage{todonotes}
\usepackage{ulem}
\newcommand{\im}{\mathrm{i}}
\newcommand{\e}{\textrm{e}}
\newcommand{\sgn}{\textrm{sgn}}

\graphicspath{{figures/}{./figures/}{.}}
\begin{document}

\title{Dimerization-induced topological superconductivity in a Rashba nanowire}

\author{Aksel Kobia\l{}ka}
\email[e-mail:]{akob@kft.umcs.lublin.pl}
\affiliation{Institute of Physics, M. Curie-Sk\l{}odowska University, 
             20-031 Lublin, Poland}

\author{Nicholas Sedlmayr}
\email[e-mail:]{sedlmayr@umcs.pl}
\affiliation{Institute of Physics, M. Curie-Sk\l{}odowska University, 
             20-031 Lublin, Poland}

\author{Maciej M. Ma\'ska}
\email[e-mail:]{maciej.maska@pwr.edu.pl}
\affiliation{Department of Theoretical Physics, Wroc{\l}aw University of Science and Technology, 
            50-370 Wroc{\l}aw, Poland}

\author{Tadeusz Doma\'nski}
\email[e-mail:]{doman@kft.umcs.lublin.pl}
\affiliation{Institute of Physics, M. Curie-Sk\l{}odowska University, 
             20-031 Lublin, Poland}

\date{\today}

\begin{abstract}
We analyze the influence of dimerization on the topological phases of a Rashba nanowire proximitized to a superconducting substrate. We find that periodic alternations of the hopping integral and spin-orbit coupling can lead to band inversion, inducing a transition to the topologically nontrivial superconducting phase that hosts Majorana zero-energy modes. This ``dimerization-induced topological superconductivity'' completely repels the topological phase of the uniform nanowire, whenever they happen to overlap. We provide an analytical justification for this puzzling behavior based on symmetry and parity considerations, and discuss feasible spectroscopic methods for its observation. We also test stability of the topological superconducting phases against electrostatic disorder.
\end{abstract}

\maketitle

\section{Introduction}
\label{sec.intro}

The topological superconducting phase of finite length one-dimensional systems with $p$-wave electron pairing enables the realization of Majorana-type quasiparticles that are immune to decoherence \cite{kitaev.01}, hence being ideal candidates for constructing stable qubits. Spectroscopic signatures of such Majorana zero-energy modes (MZMs) have been so far observed in  semiconducting nanowires proximitized to superconductors~\cite{deng.yu.12,mourik.zuo.12,das.ronen.12,finck.vanharlingen.13,deng.vaitiekenas.16,seastoft.kanne.18,lutchyn.bakkers.18,gul.zhang.18}, nanoscopic chains of magnetic atoms deposited on superconducting surfaces~\cite{nadjperge.drozdov.14,pawlak.kisiel.16,feldman.randeria.16,ruby.heinrich.17,jeon.xie.17,kim.palaciomorales.18}, lithographically fabricated nanostructures,\cite{nichele.ofarrell.17} and narrow metallic stripes embedded between two external superconductors differing in phase\cite{Fornieri-2019,Ren-2019}.

Electron pairing of these one-dimensional systems  is driven via the 
proximity effect whereas the topological phase originates either (a) from the spin-orbit coupling (SOC) combined with a sufficiently strong Zeeman 
field~\cite{sato.fujimoto.09,sato.takahashi.09,sato.takahashi.10,Lutchyn2010,oreg.refael.10} or (b) from spiral magnetic textures \cite{choy.11,Martin.Morpurgo.12,Kjaergaard2012,Bernevig2013,Simon2013,Pientka2013,klinovaja.stano.13,Vazifeh.Franz.13,Paaske2016,BlackSchaffer-2019}.
In both cases MZMs are localized on the most peripheral sites of such nanowires or nanochains~\cite{potter.lee.10,Potter2011,Sedlmayr2015,Sedlmayr2016,ptok.kobialka.17,kobialka.ptok.18.plaq,kobialka.piekarz.20} or separated by an artificial barrier~\cite{kobialka.ptok.18.ring}. Their robustness against various types of {\it perturbations} has been extensively explored, considering e.g.\  internal disorder 
\cite{Brouwer_2011,DeGottardi_2013,Hui_2015,Hoffman_2016,Hegde_2016,Pekerten_2017,maska.gorczyca.17}, disordered superconducting substrates \cite{Cole_2016}, noise \cite{Hu_2015}, inhomogeneous spin-orbit coupling \cite{Klinovaja2015},  thermal fluctuations \cite{klinovaja.stano.13,Simon2015,Scalettar2015,maska_etal_2019}, reorientation of the magnetic field~\cite{kiczek.ptok.17,Kaladzhyan2017a}, and correlations \cite{Mierzejewski_2018,Monthus_2018}.

Here, we consider a \textit{stress test} for  topological superconductivity in the Rashba nanowire that might be encountered due to dimerization. The seminal papers by Su, Schrieffer, and Heeger (SSH)  \cite{Su_1979,Heeger_1988} has firmly established that dimerization itself can induce a topological insulating phase in one-dimensional fermion systems. Interplay between dimerization and superconductivity would be currently of great importance because of its potential effect on the Majorana quasiparticles. Some aspects of the Kitaev combined with SSH scenarios have been recently addressed in Refs.~\onlinecite{Wakatsuki_2014,Wang_2017,Ezawa_2017,Chitov_2018,Yu_2019,Hua_2019}. To the best of our knowledge, however, any  systematic study of more realistic topologically superconducting nanowires is missing. For this reason we analyze here the role played by alternations of the hopping integral in a Rashba nanowire proximitized to an isotropic superconductor.

This paper is organized as follows. In Sec.~\ref{sec.model} we introduce the microscopic model describing the dimerized nanowire in presence of the Rashba and Zeeman terms that are crucial for inducing the topological superconductivity. In Sec.~\ref{sec.transition} we present the topological phase originating solely from the dimerization, and discuss its spectroscopic signatures such as the emerging Majorana quasiparticles.  In Sec.~\ref{sec_disorder} we address the electrostatic disorder and its influence on the topological phases. Sec.~\ref{sec.sum} summarizes our results and gives a brief outlook.

\section{Formulation of the problem}
\label{sec.model}

\begin{figure}
\includegraphics[width=\linewidth]{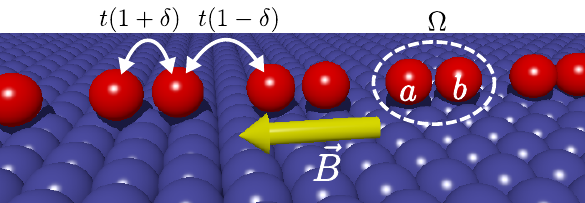}
\caption{
Schematic of the dimerized nanowire (red) deposited on the superconducting substrate (blue). Modulation of the hopping integral, $t(1\pm \delta)$, is related to shifts in the positions between neighboring $a$ and $b$ atoms forming a 2-site \textit{unit cell} $\Omega$. The (yellow) arrow shows the direction of the applied magnetic field $\vec{B}$.
}
\label{fig.scheme}
\end{figure}

We consider a semiconducting nanowire deposited on a superconductor with an alternating set of strong and weak bonds (Fig.~\ref{fig.scheme}). Modulation $\delta$ of the hopping integral and spin-orbit coupling can originate either from a mismatch of the lattice constants or due to misalignment of the nanowire with respect to the main crystallographic axes of the superconducting substrate. Neighboring atoms of the nanowire (denoted by $a$ and $b$) are not equidistant, so formally the \textit{unit cell} ${\Omega}$ comprises two sites.

The Hamiltonian of our setup, $\mathcal{H} = \mathcal{H}_{0}+\mathcal{H}_{\rm so} + \mathcal{H}_{\rm prox}$, consists first of the single-particle term 
\begin{eqnarray}
\mathcal{H}_{0} &=&-t \sum_{i,\sigma}  \left[(1+ \delta)c_{i a \sigma}^{\dagger} c_{i b \sigma}  
+ (1- \delta) c_{i a\sigma}^{\dagger} c_{i-1 b \sigma} +\mbox{\rm H.c.} \right]  \nonumber \\
&&- \sum_{{\Omega} = a,b} \sum_{i,\sigma}  \left( \mu + \sigma^{z}_{\sigma\sigma}h \right) 
c_{i{\Omega}\sigma}^{\dagger} c_{i{\Omega}\sigma}\,,
\label{,SSH_part}
\end{eqnarray}
describing electrons moving along the periodically deformed nanowire. The second quantization operators $c_{i{\Omega}\sigma}^{\dagger}$ ($c_{i{\Omega}\sigma}$) create (annihilate) an electron with spin $\sigma$ at site $\Omega=a$ or $b$ of the {\it i}-th \textit{unit cell}, $\mu$ is the chemical potential and $h$ stands for the Zeeman shift induced by the magnetic field. The hopping integral $t(1 \pm \delta)$ between the nearest-neighbor sites periodically varies with a relative amplitude $\delta$. The same modulation is also imposed in the spin-orbit Rashba term
\begin{eqnarray}
\mathcal{H}_{\rm so} &= &- \im\lambda \sum_{i \sigma \sigma'} \left[ (1+ \delta) c_{i a \sigma}^{\dagger} 
\sigma^{y}_{\sigma\sigma'} c_{i b \sigma'} \right. \label{Rashba_term} \\ \nonumber 
&&+ \left. (1- \delta)c_{i a \sigma}^{\dagger}\sigma^{y}_{\sigma\sigma'} 
c_{i-1 b \sigma'} \right] + \mbox{\rm H.c.} \,,
\end{eqnarray}
where $\sigma^{x,y,z}$ are the Pauli matrices. The last part $\mathcal{H}_{\rm prox}$ accounts 
for the proximity induced on-site electron pairing. For simplicity we describe it by the BCS-like term 
\begin{eqnarray}
\mathcal{H}_{\rm prox} =   \sum_{i} \sum_{{\Omega}=a,b}  \left( \Delta 
c_{i{\Omega}\uparrow}^{\dagger} c_{i{\Omega}\downarrow}^{\dagger}  + 
\Delta^{\ast} c_{i{\Omega}\downarrow}c_{i{\Omega}\uparrow} \right)\,.
\end{eqnarray}

Previous considerations of the uniform Rashba nanowire have established that the topologically non-trivial superconducting phase is realized for magnetic fields obeying the constraint~\cite{sato.fujimoto.09}
\begin{eqnarray}
\label{eq.relation} \sqrt{ \left( 2 t - \mu \right)^{2} + | \Delta |^{2} } < h < \sqrt{ \left( 2 t + \mu \right)^{2} + | \Delta |^{2} } \,.
\end{eqnarray}
The topological phase transition occurs when the quasiparticle spectrum closes and reopens the soft gap~\cite{moore.balents.07}. In Sec.~\ref{sec.transition} we shall revisit this criterion in presence of dimerization $\delta$ and  determine the topological phase diagram with respect to the model parameters $h$, $\lambda$, $\delta$, $\Delta$, and $\mu$. 

\subsection{Formalism} 

The Hamiltonian $\mathcal{H}$ can be recast in the Nambu basis
\begin{equation}
\Psi_i = \left(c_{ia\uparrow},c_{ib\uparrow},c_{ia\downarrow},
c_{ib\downarrow}, c^\dagger_{ia\downarrow},c^\dagger_{ib\downarrow},
-c^\dagger_{ia\uparrow},-c^\dagger_{ib\uparrow}\right)^{T} 
\label{Nambu_basis}
\end{equation}
using the Bogoliubov-de~Gennes procedure. 
We then diagonalize the matrix $H_{ij}$
defined via $\mathcal{H}=\frac{1}{2}\sum_{i,j}\Psi_{i}^\dagger H_{ij} \Psi_{j}$. Its Fourier transform $\mathcal{H}_k$ takes the form
\begin{equation}
    H_k=-h\sigma^z-\mu\tau^z-t\gamma_k^+\tau^z-- i \lambda \gamma^{-}_{k} \sigma^y \tau^z+\Delta\tau^x\,,
\end{equation}
where $\tau^{x,y,z}$ are Pauli matrices acting within the particle-hole subspace, and we have assumed (without loss of generality) that $\Delta$ is real. We have additionally introduced the matrices acting in the sublattice space
\begin{equation}
    \gamma_k^\pm=\left(\begin{array}{cc}
    0&(1+\delta)\pm(1-\delta)\e^{\im k}\\
    \pm(1+\delta)+(1-\delta)\e^{-\im k}&0
\end{array}\right)\,.
\end{equation}
By convention identity matrices are not explicitly shown and a tensor product over the matrices is implied.

\subsection{Experimentally accessible observables}

In specific numerical computations we have considered a finite length nanowire consisting of $200$ sites and used $\Delta= 0.2t$, $\lambda=0.15t$, and $h=0.3t$ (unless stated otherwise). Typical values of the hopping integral between the nearest neighbor atoms on superconducting surface are $t\sim 10$ meV \cite{Wiesendanger-2012,Yazdani-2013}, whereas their spacing varies between 0.3 and 0.6 nm, see Table 1 in Ref.~\cite{Klinovaja-2013}.
The eigenvalues $\varepsilon_{n}$ and eigenvectors in Nambu space $\left(u^n_{ia\uparrow}, u^n_{ib\uparrow}, u^n_{ia\downarrow},
u^n_{ib\downarrow}, v^n_{ia\downarrow}, v^n_{ib\downarrow}, v^n_{ia\uparrow}, v^n_{ib\uparrow}\right)^{T}$ (see Eq.~\eqref{Nambu_basis}) are determined by numerical diagonalization, from which we construct the local density of states 
\begin{equation}
    \rho_{i{\Omega}} ( \omega ) 
    = \sum_{\sigma n} |u^n_{i \Omega \sigma}|^2\delta(\omega-\varepsilon_n)+|v^n_{i \Omega  \sigma}|^2\delta(\omega+\varepsilon_n)\,.
\end{equation}
This local density of states (LDOS) is measurable by scanning tunneling microscopy (STM)~\cite{wiesendanger.09} and, at low temperatures, is equivalent to the differential conductance $G_{i{\Omega}} = dI_{i{\Omega}}(V)/dV$ of the tunneling current $I_{i{\Omega}}(V)$ induced by a voltage $V$~\cite{figgins.morr.10}. In special cases it is useful to inspect the total density of states (DOS) obtained from the summation $\rho(\omega)=\sum_{i{\Omega}} \rho_{i{\Omega}} ( \omega )$. Since our numerical solution is obtained for a finite size nanowire we shall illustrate the resulting spectra replacing the Dirac delta functions by Gaussian distributions, $\delta ( \omega-\varepsilon_{n} ) = \frac{1}{\sigma \sqrt{2 \pi}} \exp{\left( \frac{-(\omega-\varepsilon_{n})^{2}}{2 \sigma^{2}}\right) }$, with a small broadening $\sigma= 0.0035t$.

\section{Topological superconductivity}
\label{sec.transition}

\begin{figure}[b]
\includegraphics[width=0.49\linewidth,keepaspectratio]{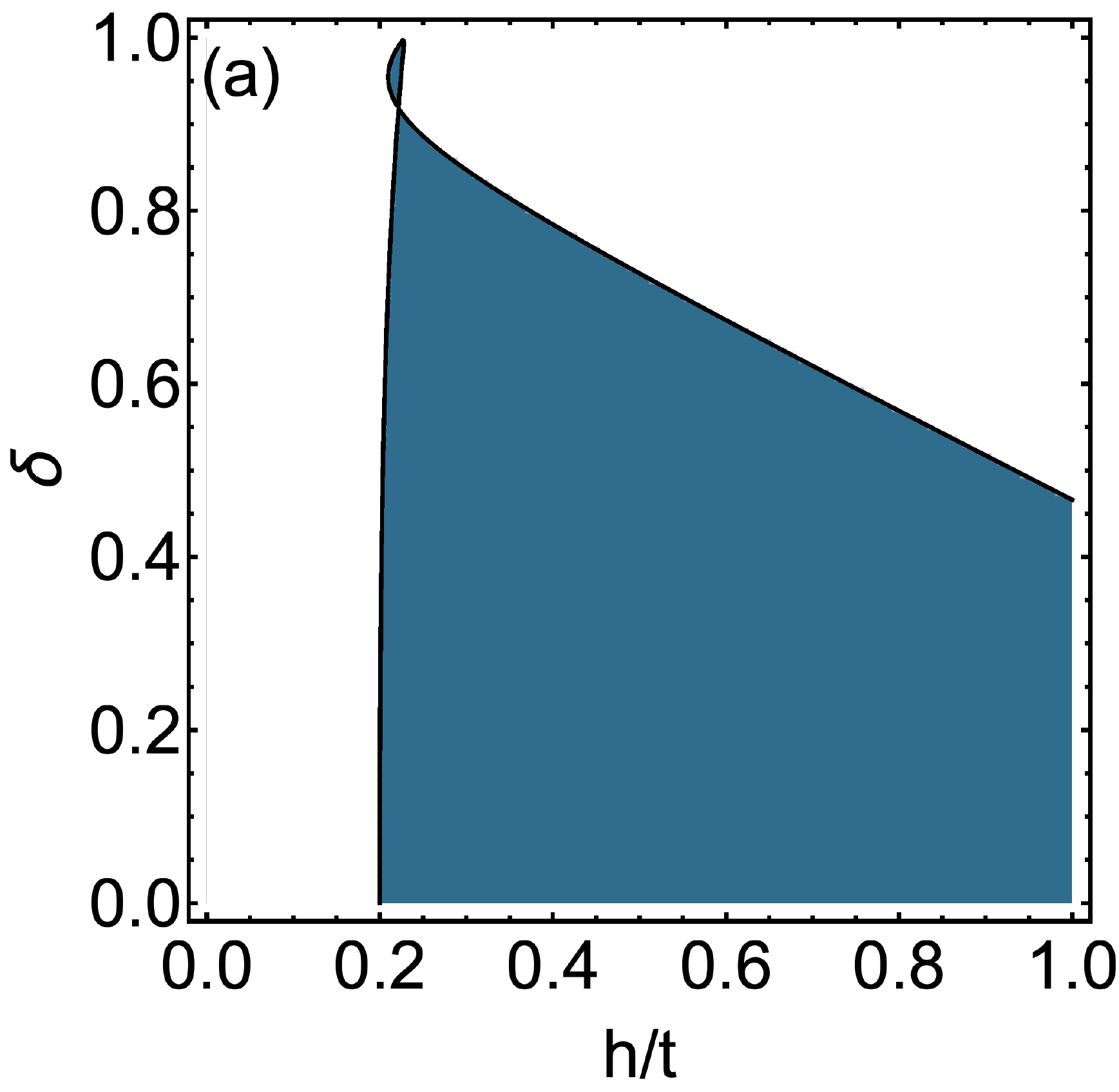}
\includegraphics[width=0.49\linewidth,keepaspectratio]{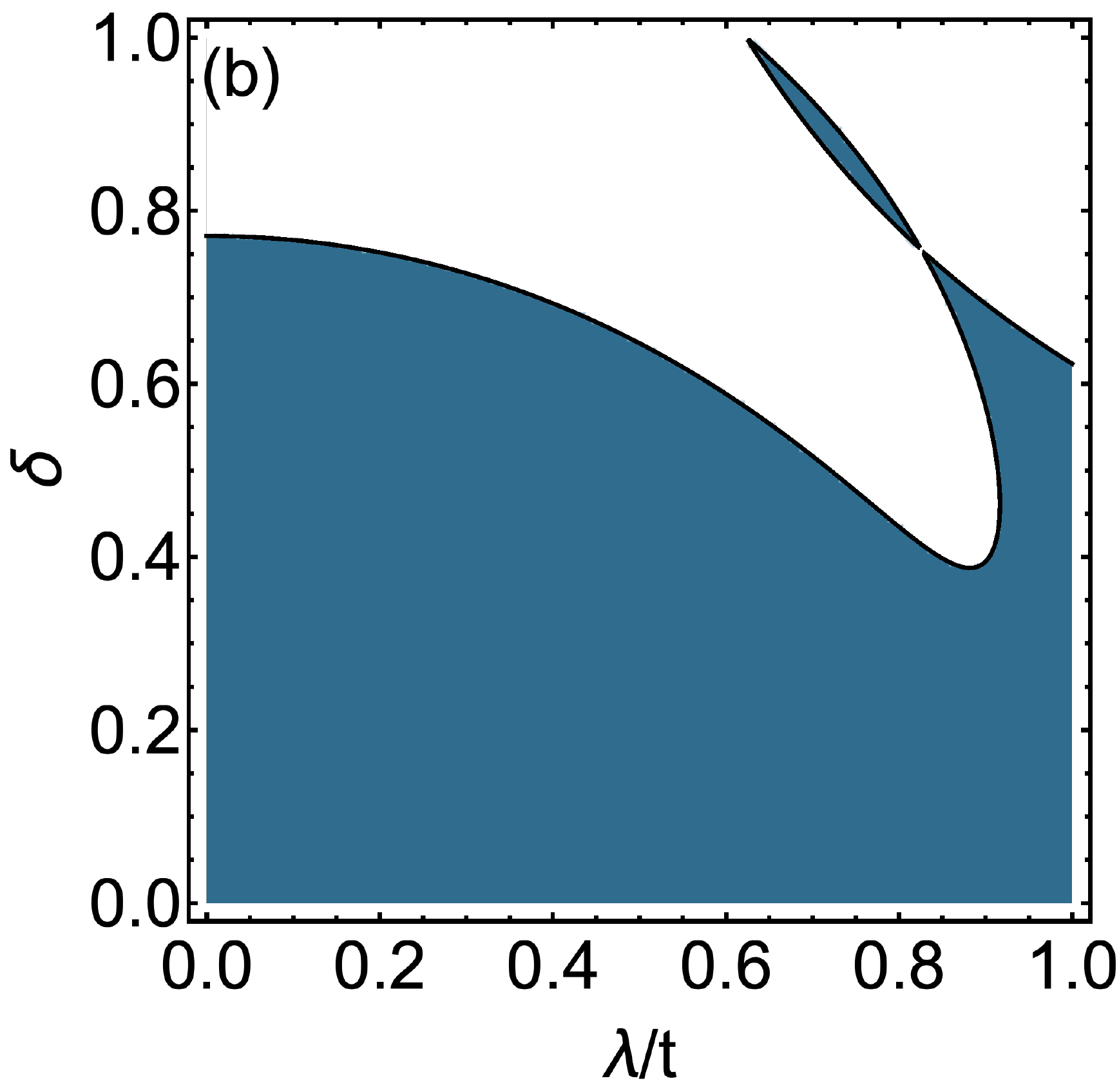}
\caption{
Topological phase diagram of the dimerized nanowire. Panel (a) shows the phase diagram with respect to the magnetic field $h$ and the hopping integral modulation $\delta$, obtained for $\lambda=0.3t$; panel (b) with respect to the spin-orbit coupling $\lambda$ and the hopping integral modulation $\delta$, for $h = 0.3t$. The dark (blue) regions have a parity of $(-1)^\nu=-1$ and are topologically non-trivial, the white regions are topologically trivial. Gap closings at $k=0$ are marked with solid black lines and at $k=\pi$ with dashed black lines. 
Parameters used for both plots are: $\Delta = 0.2 t$, $\mu = -2t$.
} 
\label{fig.hvdt}
\end{figure}

\begin{figure}
\includegraphics[width=0.95\linewidth,keepaspectratio]{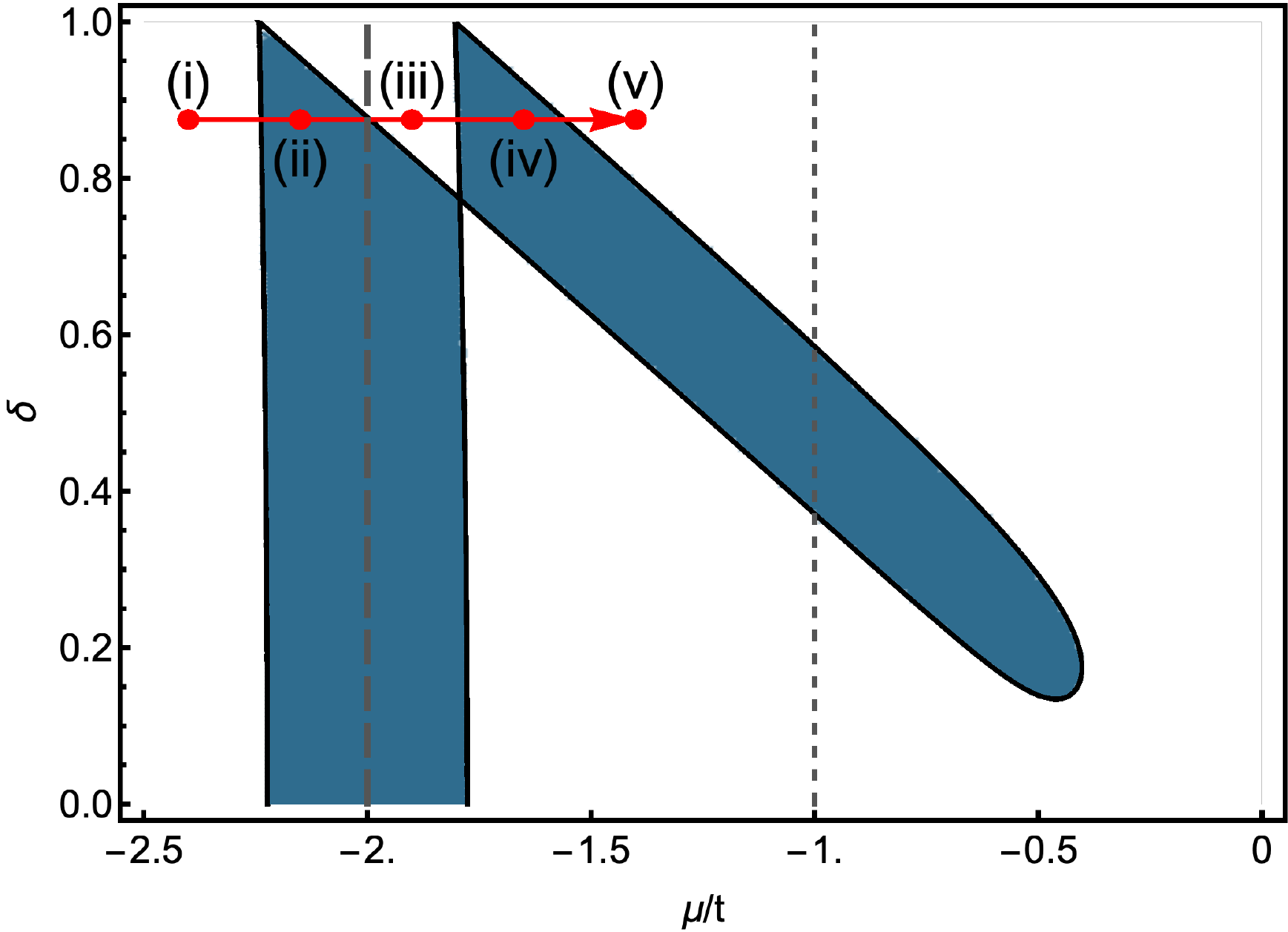}
\caption{
Topological phase diagram of the dimerized nanowire with respect to the chemical potential $\mu$ and the modulation of the hopping integral $\delta$. The dark (blue) regions have a parity of $(-1)^\nu=-1$ and are topologically non-trivial, the white regions are topologically trivial. The dimerization introduces a new region of topology at smaller chemical potential than in the homogeneous case. The band structures at the points (i)-(v), demonstrating the band inversion, are shown in Fig.~\ref{fig.bands}. The dashed lines correspond to the calculations of the DOS and LDOS, see Figs.~\ref{fig.dos1} to \ref{fig.ldos2}. 
Results calculated for $\Delta = 0.2 t$, $h=0.3t$ and $\lambda=0.15t$.
}
\label{fig.mudt}
\end{figure}

The dimerized nanowire still retains the symmetries of the homogeneous wire, namely particle-hole $[H,\mathcal{C}]_+=0$, where $\mathcal{C}= \sigma^{y} \tau^{y} \hat{K}$, and a ``time-reversal'' symmetry $[H,\mathcal{T}]_-=0$, where $\mathcal{T}=\sigma^x\hat{K}$ and $\mathcal{T}^2=1$. $\hat{K}$ is the complex conjugation operator. Hence the Hamiltonian also possesses their composite symmetry $\mathcal{T}\mathcal{C}$, often referred to as chiral symmetry. The Hamiltonian is therefore in the BDI class \cite{Ryu2010} and has a $\mathbb{Z}$ topological index, the winding number $\nu$ . However, for this particular Hamiltonian,we find that this index does not obtain magnitudes larger than 1, and hence all interesting information from this index can also be contained in its parity $(-1)^\nu$. This is relatively straightforward to calculate, using either the Pfaffian \cite{kitaev.01}, scattering matrices \cite{akhmerov.11,fulga.11}, or a suitable parity operator \cite{Sato2009b,sato.fujimoto.09,sato.takahashi.09}. Here we will focus on the last option, as this also allows us to understand the phase diagrams inferred from band inversions.

The index $\nu$ can be related directly to an appropriately defined parity of the negative energy bands at the time reversal invariant momenta $\{\Gamma_{1},\Gamma_{2}\}=\{0,\pi\}.$ The parity operator $\mathcal{P}$ must satisfy $[\mathcal{P},\mathcal{C}]_+=0$ and $[\mathcal{P},H_{\Gamma_i}]_-=0$. In that case the eigenstates $\left| n,k\right>$ at the time-reversal invariant points are eigenstates also of the parity operator and have a definite parity $\Pi_{n,\Gamma_i}\equiv\langle n,\Gamma_i|\mathcal{P}|n,\Gamma_i\rangle=\pm1$. One can then demonstrate that \cite{Sato2009b,Dutreix2017}
\begin{equation}
    (-1)^\nu=\prod_{\varepsilon_{n,\Gamma_i}<0}\Pi_{n,\Gamma_i}\,.\label{eq.z2}
\end{equation}
Calculation of the topological phase is therefore reduced to finding a suitable parity operator. Following the methods of Refs.~\onlinecite{Sedlmayr2015b,Sedlmayr2016,Dutreix2017,Sedlmayr2017} we find $\mathcal{P}=\lambda^{x} \sigma^{z}$, where $\lambda^{x,y,z}$ are Pauli matrices acting in the sublattice subspace.

Finally one finds
\begin{eqnarray}
    (-1)^\nu=\sgn
    \bigg[\left(h^2-\mu^2\right)^2+\left(4t^2+4\lambda^{2}\delta^{2}+\Delta^2\right)^2\nonumber\\
    -2\mu^2\left(4t^2+4\lambda^{2}\delta^{2}-\Delta^2\right)-2h^2\left(4t^2-4\lambda^{2}\delta^{2}+\Delta^2\right)\bigg]
    \nonumber\\
    \times\sgn\bigg[\left(h^2-\mu^2\right)^2+\left(4\lambda^{2}+4t^2\delta^{2}+\Delta^2\right)^2 \label{z2eqn_index}\qquad\\\nonumber
    -2\mu^2\left(4\lambda^{2}+4t^2\delta^{2}-\Delta^2\right)+2h^2\left(4\lambda^{2}-4t^2\delta^2-\Delta^2\right)\bigg]
    \,.
\end{eqnarray}
The first terms change sign when the gap closes at $k=0$, and the second when it closes at $k=\pi$. 
These two conditions are marked separately on Figs.~\ref{fig.hvdt} -- \ref{fig.muh}. The gap closing lines separating topologically trivial and non-trivial regions are given in Appendix~\ref{appendix}, as well as the expression in the limit $h,\mu\to\infty$. For $\delta=0$ one finds
\begin{eqnarray}
    (-1)^\nu&=&\sgn
    \left[\left(4t^2-h^2+\Delta^2\right)^2\right.
    \\\nonumber&&\qquad
    \left.-2\mu^2\left(4t^2+h^2-\Delta^2\right)+\mu^4\right]\,,
\end{eqnarray}
which reproduces the well-known result for a homogeneous wire. In that limit the second term in \eqref{z2eqn_index} becomes positive definite and no longer contributes. As for some quasi-one dimensional wires \cite{Sedlmayr2016} and hexagonal lattices \cite{Dutreix2017,Sedlmayr2017}, which are related to the dimerized wire, the topological phase now depends explicitly on the strength of the spin-orbit coupling $\lambda$. The additional conditions for topology also indicate, as we shall see, that there are new topologically non-trivial phases appearing.


\begin{figure}
\includegraphics[width=0.49\linewidth,keepaspectratio]{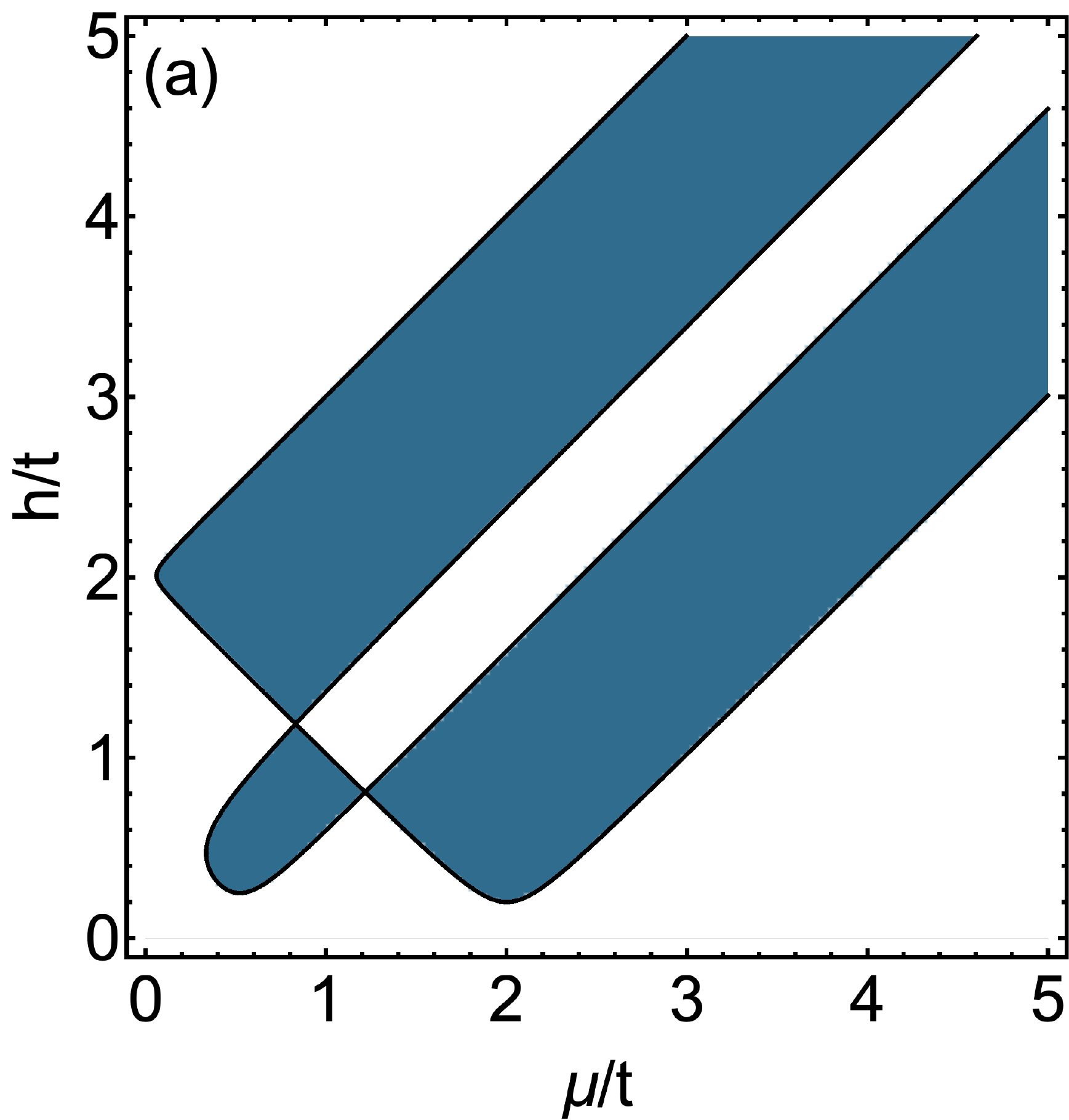}
\includegraphics[width=0.49\linewidth,keepaspectratio]{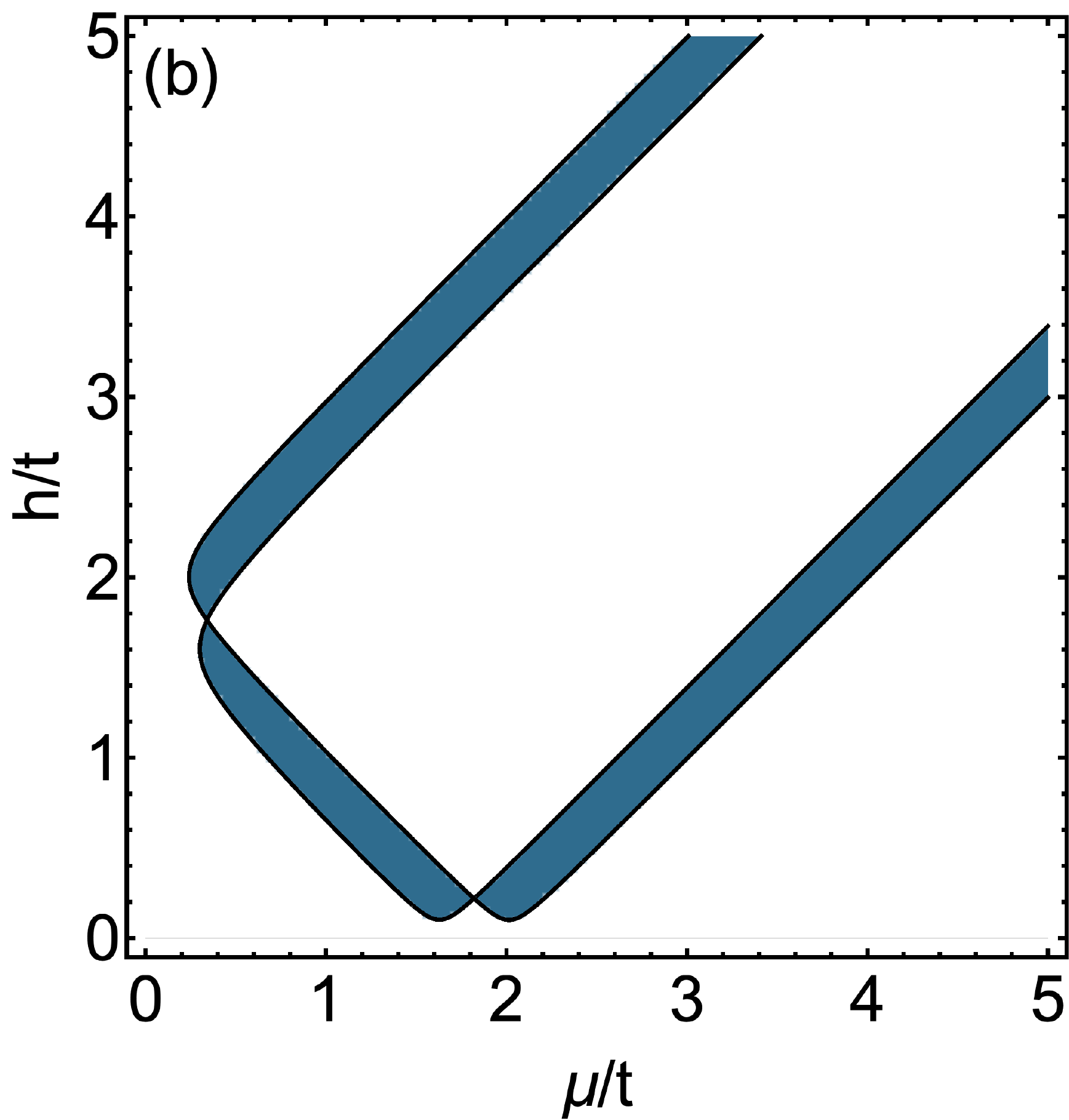}
\caption{
Topological phase diagram of the dimerized nanowire 
with respect to the chemical potential $\mu$ and the magnetic field $h$. The dark (blue) regions have a parity of $(-1)^\nu=-1$ and are topologically non-trivial, the white regions are topologically trivial. Panel (a) is for $\delta=0.2$ and panel (b) for $\delta=0.8$. The dimerization destroys a part of the topological phase and at a critical value there is no topologically non-trivial phase left, as seen also in Figs.~\ref{fig.hvdt} and \ref{fig.mudt}. Gap closings at $k=0$ are marked with solid black lines and at $k=\pi$ with dashed black lines.
Both results were obtained for $\Delta = 0.2 t$, $h = 0.3t$ and $\lambda = 0.15t$
}
\label{fig.muh}
\end{figure}

Examples of the phase diagrams  are displayed in Figs.~\ref{fig.hvdt}--\ref{fig.muh}. Stability of the topological superconducting state of the proximitized Rashba nanowire is very sensitive to magnetic field. Fig.~\ref{fig.hvdt}(a) depicts the phase diagram with respect to the applied magnetic field $h$ and the hopping modulation $\delta$. The lowest critical field is $h \simeq 0.2t$ and it is rather unaffected by dimerization. Contrary to this, the upper critical field is considerably suppressed by dimerization. The lower and upper critical magnetic fields merge at sufficiently strong dimerization ($\delta \approx 0.95$).

Furthermore, we would like to emphasize the appearance of the additional topological phase induced solely by the dimerization as can be seen in Fig.~\ref{fig.mudt}. Such additional topological phase forms away from the usual topological phase of the uniform nanowire existing around $\mu=-2t$. This dimerization induced topologically non-trivial phase nonetheless still requires spin-orbit coupling to be present. 

\subsection{Band inversion}

\begin{figure}[t]
\includegraphics[width=0.49\linewidth,keepaspectratio]{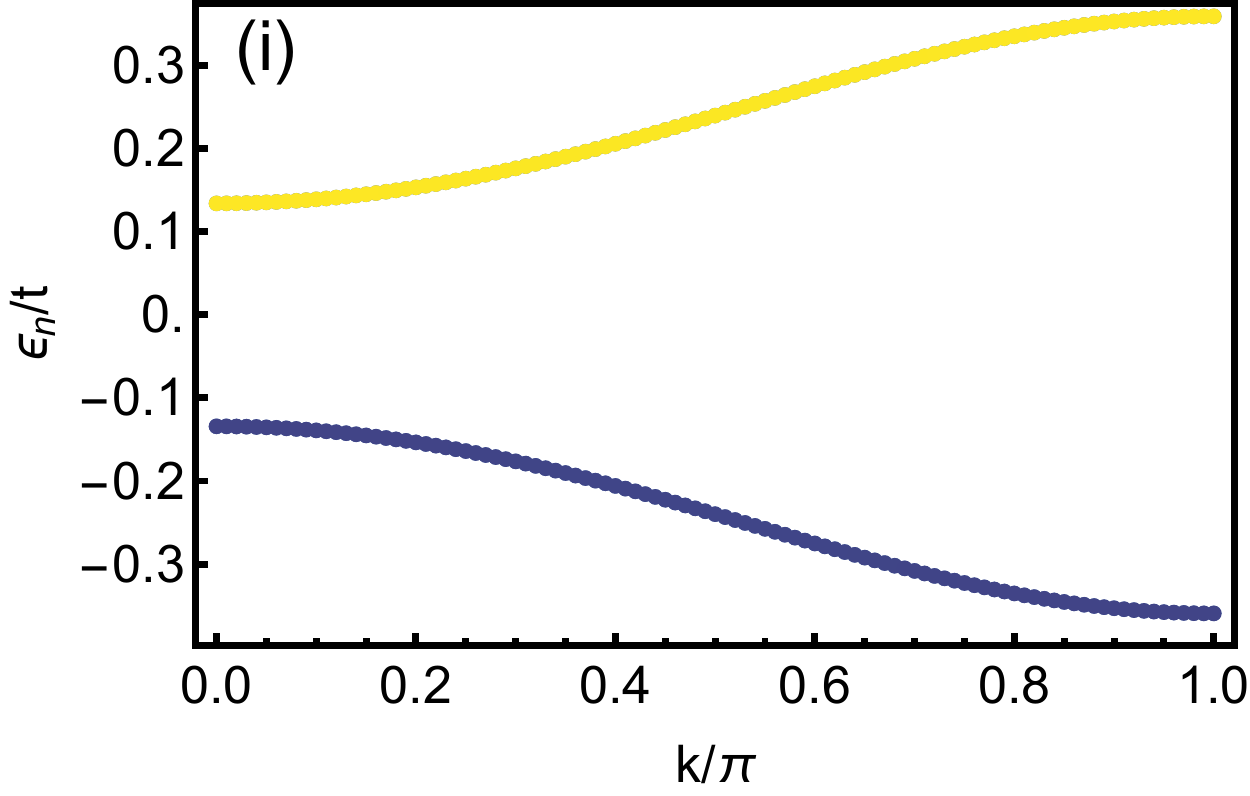}
\includegraphics[width=0.49\linewidth,keepaspectratio]{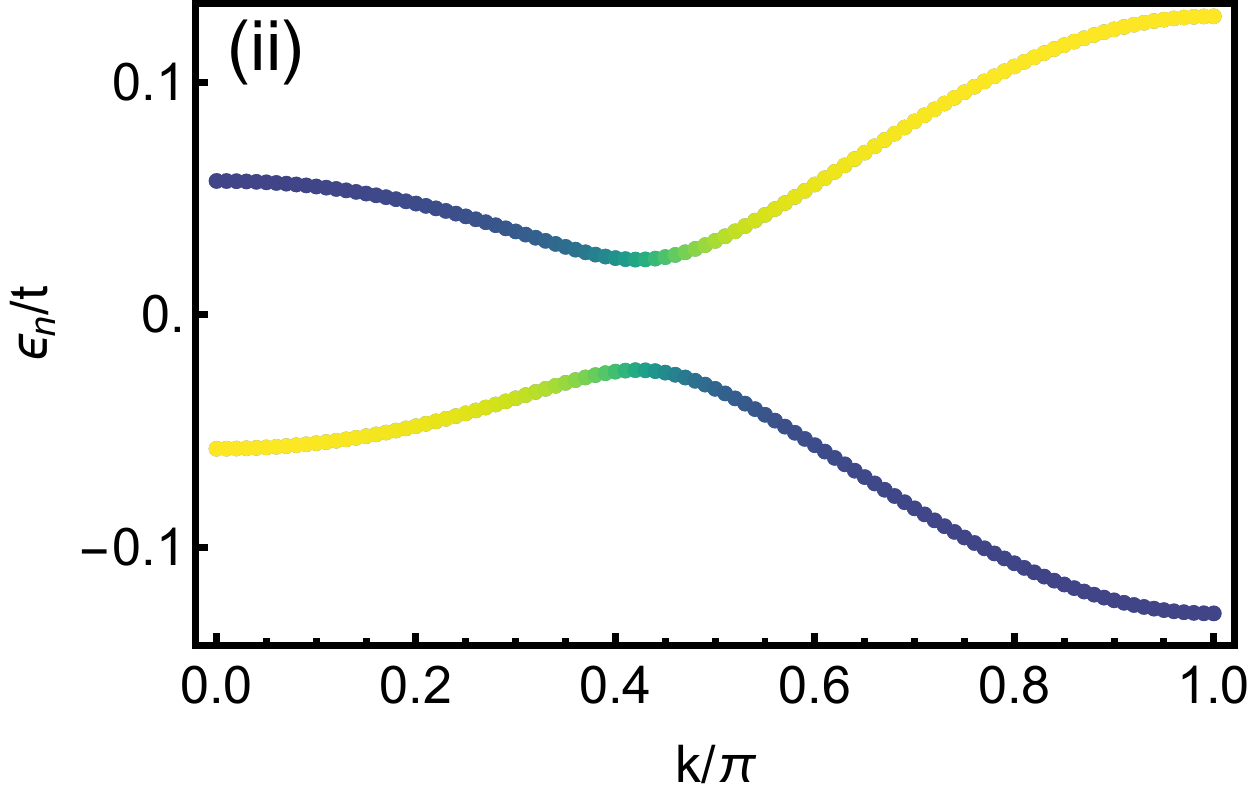}\\
\includegraphics[width=0.49\linewidth,keepaspectratio]{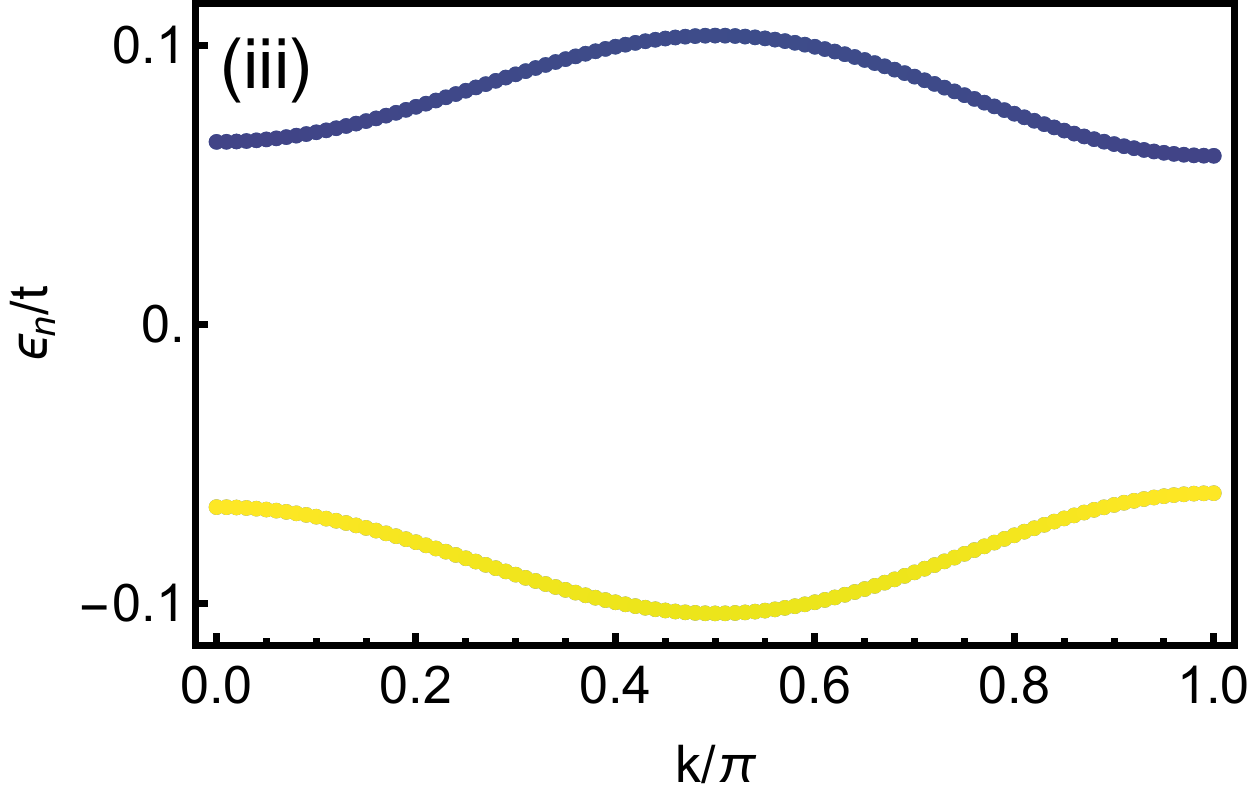}
\includegraphics[width=0.49\linewidth,keepaspectratio]{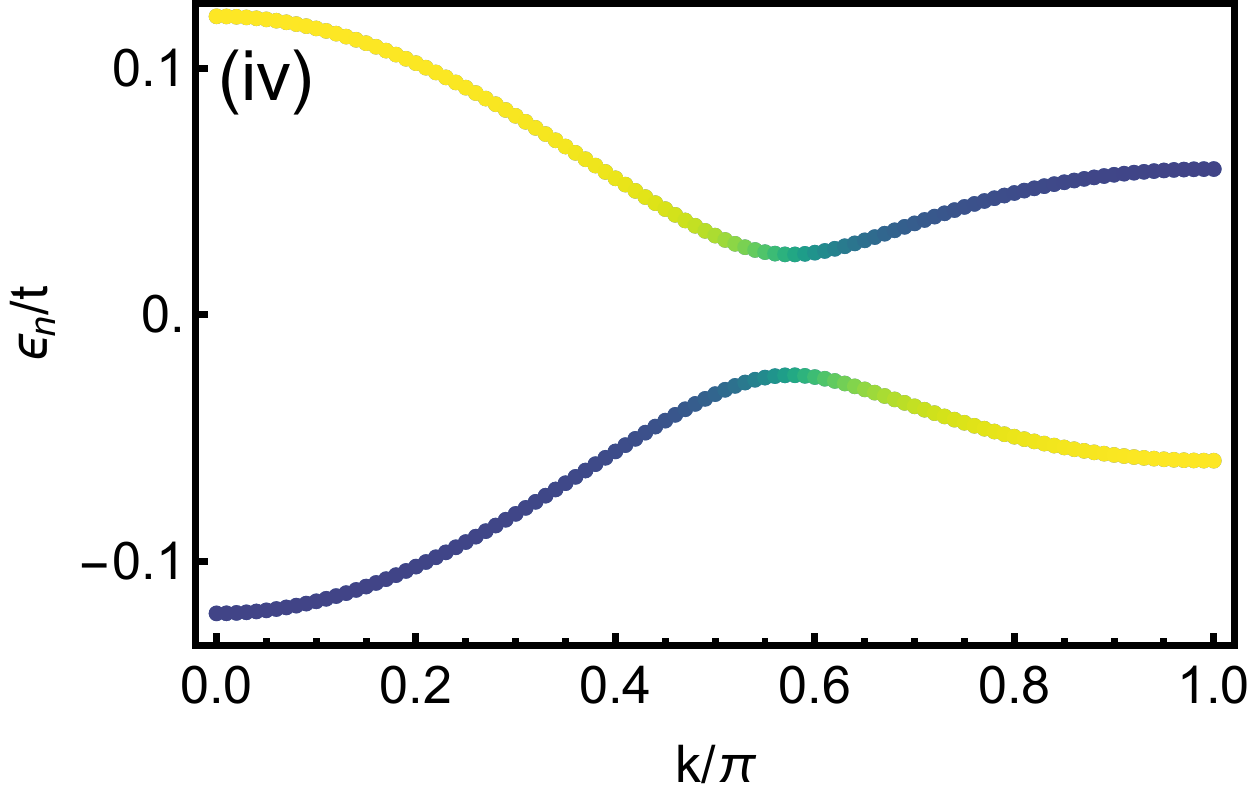}\\
\includegraphics[width=0.49\linewidth,keepaspectratio]{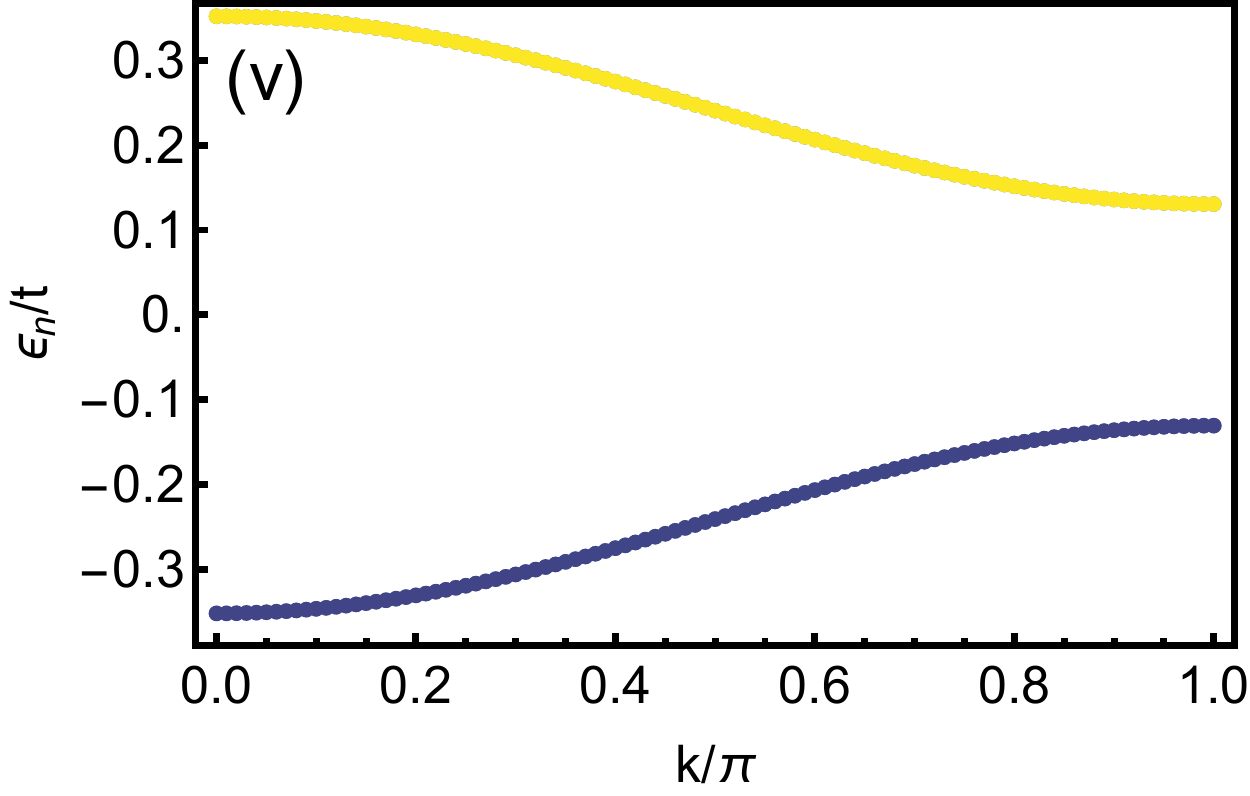}
\includegraphics[width=0.49\linewidth,keepaspectratio]{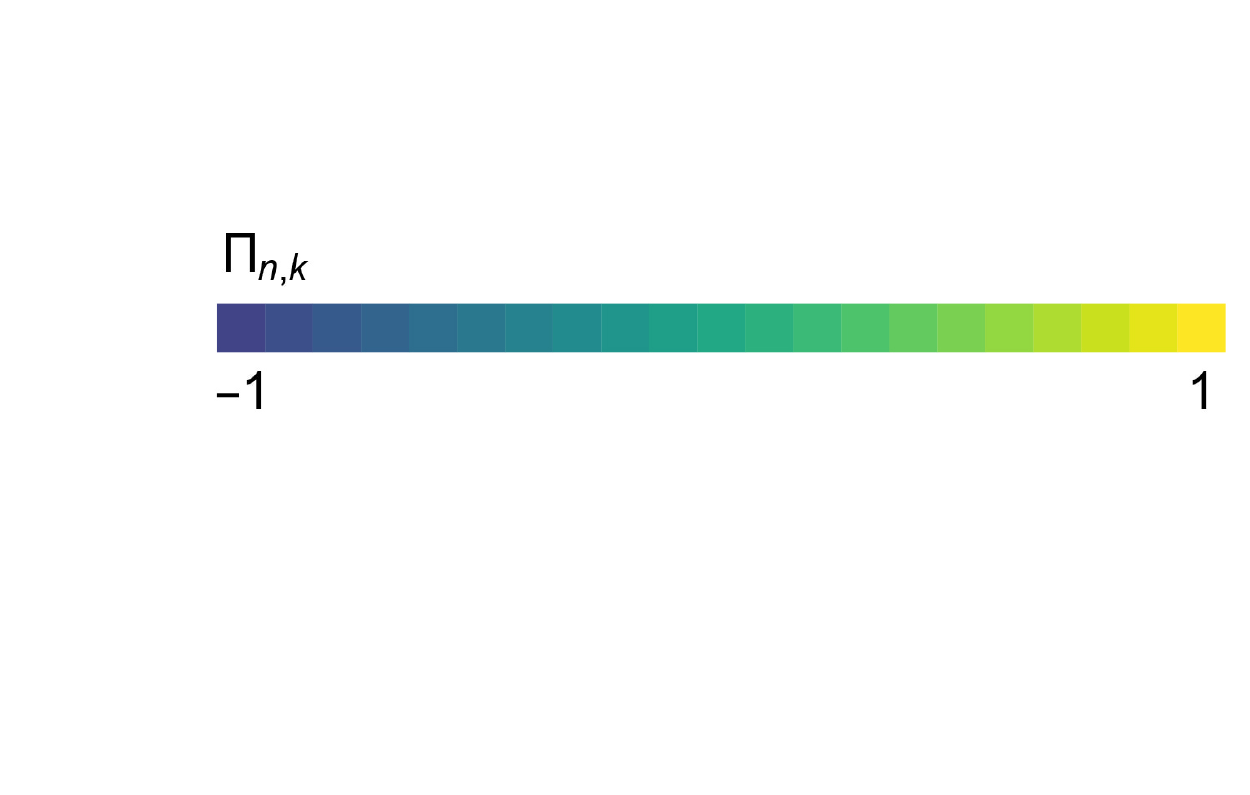}
\caption{
Band inversion demonstrating the topological phases. Shown are the lowest negative and positive energy bands between the time-reversal invariant momenta $0$ and $\pi$. Between each panel is a gap closing (at either $k=0$ or $k=\pi$) which inverts the parity of the bands (see main text for more details). Parameters in panels (i)-(v) are indicated in Fig.~\ref{fig.mudt}, with $\delta=0.875$ and the chemical potential: (i) $\mu=-2.4t$; (ii) $\mu=-2.15t$; (iii) $\mu=-1.9t$; (iv) $\mu=-1.65t$; and (v) $\mu=-1.4t$
Rest of the parameters are as follows: $\Delta = 0.2 t$, $h = 0.3t$ and $\lambda = 0.15t$.}
\label{fig.bands}
\end{figure}

Using the parity operator from which the topological index was calculated, one can demonstrate the topology by considering band inversion. One can define the parity of a band at a momentum $k$ as
\begin{equation}
    \Pi_{n,k}\equiv\langle n,k|\mathcal{P}|n,k\rangle\,.
\end{equation}
At $k=0,\pi$ the energy eigenstate is also an eigenvector of parity with eigenvalues $\pm1$. From the definition of the index Eq.~\eqref{eq.z2} it should be apparent that the system is in a topologically non-trivial phase when the parity of the negative energy bands switches an odd number of times between the time reversal invariant momenta.

We can check this explicitly for the phases shown in Fig.~\ref{fig.mudt}, see Fig.~\ref{fig.bands}. Between panels (i) and (ii) the gap closes and opens with band inversion occurring. The gap closing associated with the new topological phase re-inverts these bands and the system becomes trivial again for (iii). The subsequent gap closing and opening from (iii) to (iv) pushes the system into the new topological phase. This phase has the bands inverted along a different orientation of $k$, which is why these two phases destroy each other, becoming topologically trivial, when they cross (see Figs.~\ref{fig.mudt}--\ref{fig.muh}).

\subsection{Quasiparticle spectra}

\begin{figure}[t!]
\includegraphics[width=\linewidth,keepaspectratio]{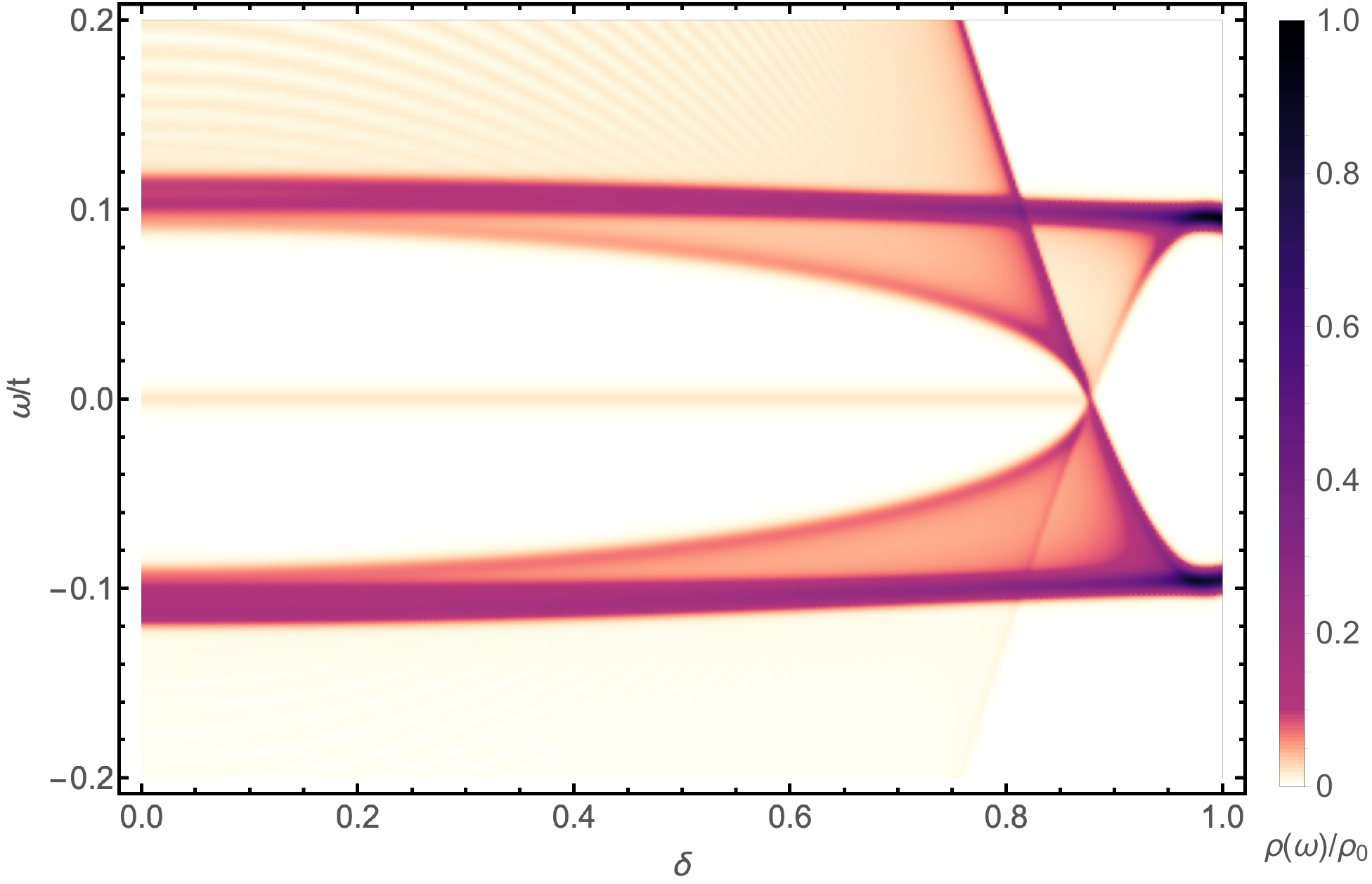}
\caption{
Evolution of the density of states $\rho(\omega)$ 
upon the modulation $\delta$ obtained for $\mu=-2t$, $h=0.3t$, $\lambda = 0.15t$, and $\Delta = 0.2t$. The density of states is scaled by $\rho_0=1.26\times 10^{4}t^{-1}$. 
\label{fig.dos1}
}
\end{figure}

Let us now inspect the evolution of the quasiparticle spectra driven by  dimerization. In Fig.~\ref{fig.dos1} we show the density of states obtained for the model parameters $\mu=-2t$, $h=0.3t$, $\lambda =0.15t$. We can notice that the soft gap gradually closes upon approaching the critical $\delta=0.87$, and the system evolves into the topologically trivial phase. Traversing this critical dimerization we clearly observe signatures of the band inversion accompanied by disappearance of the Majorana quasiparticles. Ultimately, for $\delta \rightarrow 1$ the nanowire becomes entirely dimerized, therefore its spectrum evolves to  the bonding and anti-bonding states. We have checked that for larger values of the magnetic field, the topological phase, and hence the MZMs, are destroyed at considerably lower dimerization strengths
(Fig. \ref{fig.hvdt}).

\begin{figure}
\includegraphics[width=\linewidth,keepaspectratio]{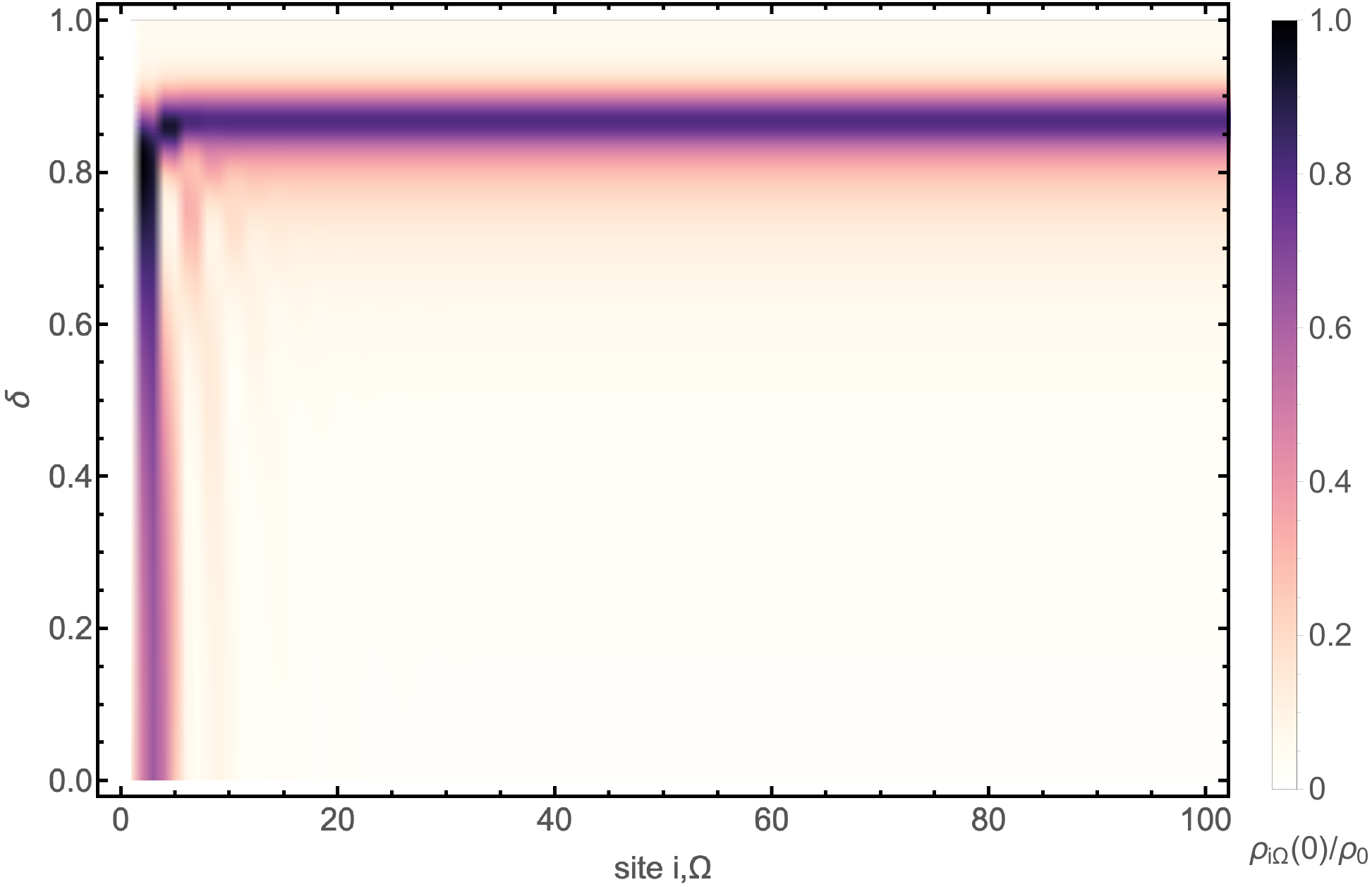}
\caption{The local density of states at zero energy $\rho_{i\Omega}(\omega)$ as a function of the hopping integral modulation $\delta$ obtained for the same model
parameters as in Fig.~\ref{fig.dos1}. The MZMs can be clearly seen at the edges of the wire, until the critical dimerization closes the gap. The normalization is $\rho_0=4.65t^{-1}$. 
Only data for the left half of the nanowire (first $100$  sites) is shown, as the nanowire is symmetric.
} 
\label{fig.ldos1}
\end{figure}

In Fig.~\ref{fig.ldos1} we illustrate the changeover of the Majarona profile driven by dimerization. For this purpose we display the LDOS at zero energy $\rho_{i{\Omega}} ( \omega\!=\!0 )$ with respect to sites $\{i,\Omega\}\in \left< 1 ; N/2 \right>$ and for varying $\delta$. The spatial profile of the MZM is rather stable for a wide range of the hopping integral modulation $\delta$. Upon approaching the critical value $\delta \approx 0.87 t$ the topological transition, caused by the band inversion, occurs. The zero-energy Majorana modes then cease to exist and merge back into the bulk states.

\begin{figure}
\includegraphics[width=\linewidth,keepaspectratio]{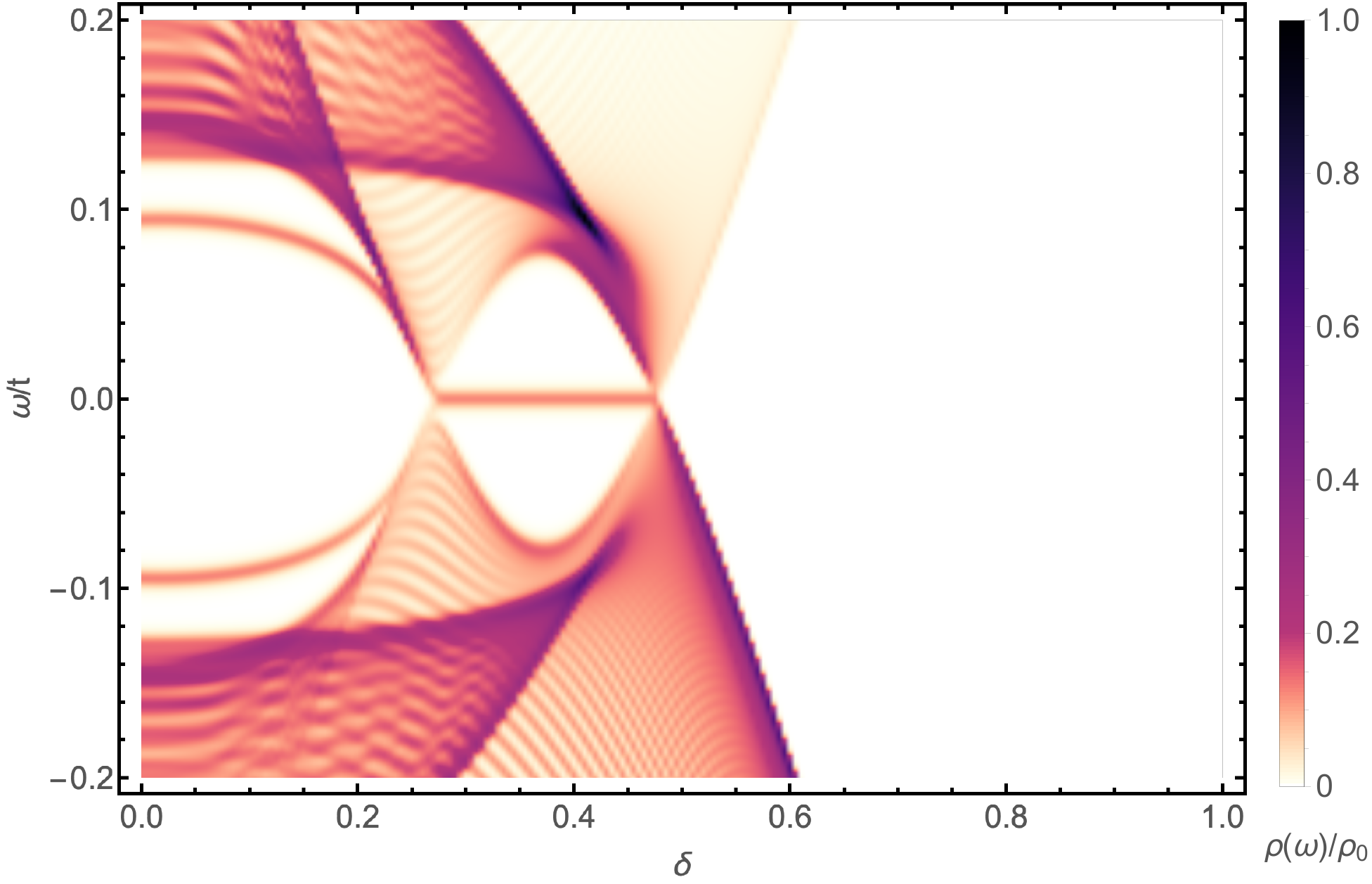}
\caption{
Evolution of the density of states $\rho(\omega)$ driven by the hopping integral 
modulation $\delta$ obtained for $\mu=-0.8t$, $h=0.3t$, $\lambda = 0.15t$, and $\Delta = 0.2t$, i.e.,~within the new topological phase induced by the dimerization. The density of states is scaled by $\rho_0=1.89\times 10^{3}t^{-1}$.
}
\label{fig.dos2}
\end{figure}

\begin{figure}
\includegraphics[width=\linewidth,keepaspectratio]{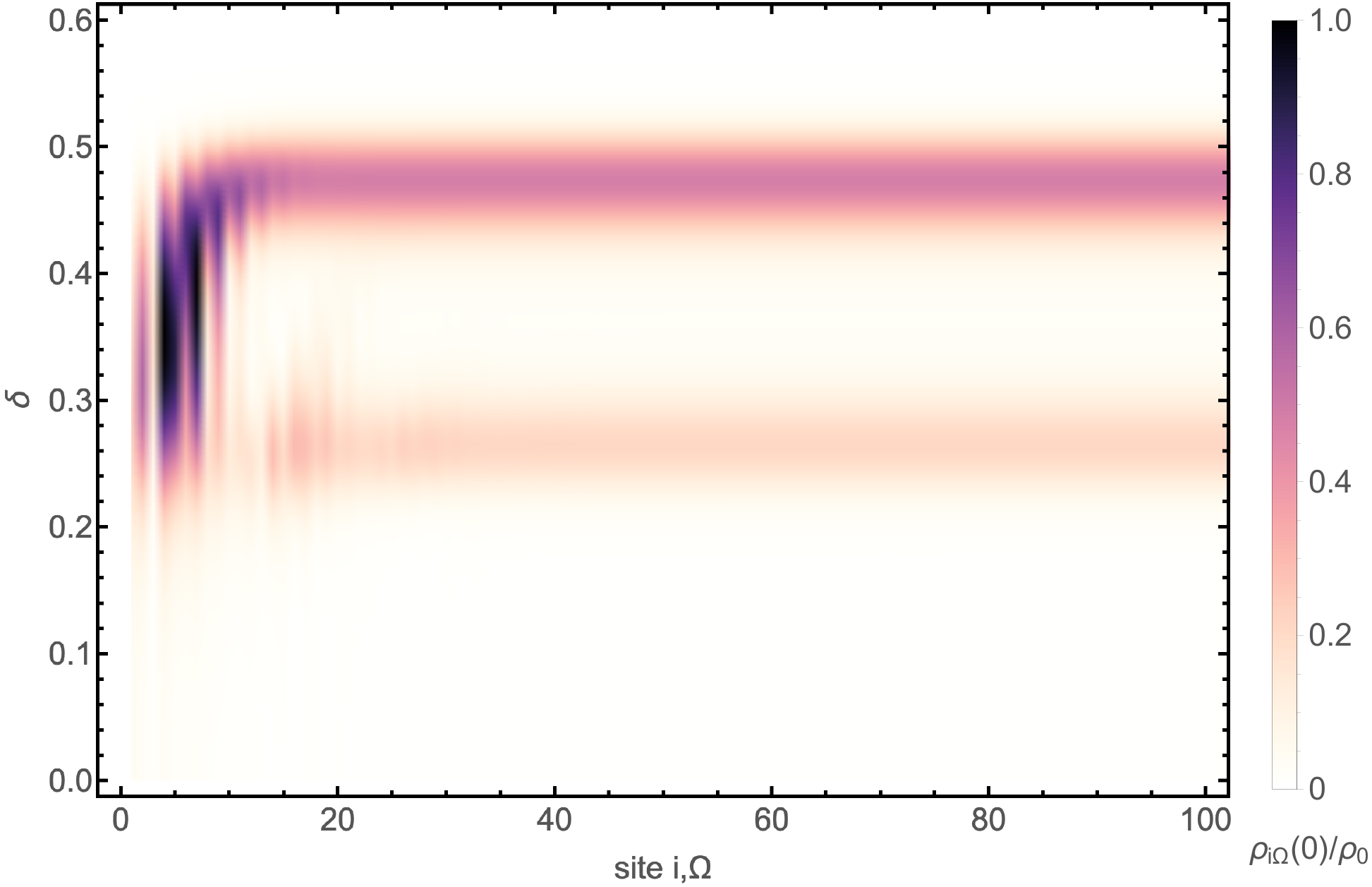}
\caption{
The local density of states at zero energy $\rho_{i\Omega}(\omega)$ as a function of the hopping integral modulation $\delta$ obtained for the same model parameters as in Fig.~\ref{fig.dos2}. The MZMs can be clearly seen at the edges of the wire, until the critical dimerization closes the gap. The normalization is $\rho_0=2.14t^{-1}$. Only data for the left half of the nanowire are shown, the right half are symmetric. The very faint edge states that can be seen for small dimerization are traces of the trivial non-zero energy subgap states which are clearly visible in Fig.~\ref{fig.dos2}.
}
\label{fig.ldos2}
\end{figure}

Figure~\ref{fig.dos2} shows the density of states obtained for $\mu=-1t$, corresponding to the topologically non-trivial phase driven by dimerization. In this case the MZMs are present over a finite dimerization regime, between the subsequent gap closing points signaling the change in topology, as can be clearly seen in Fig.~\ref{fig.ldos2}.

\section{Robustness to disorder}
\label{sec_disorder}

\begin{figure}[b]
\includegraphics[width=\linewidth,keepaspectratio]{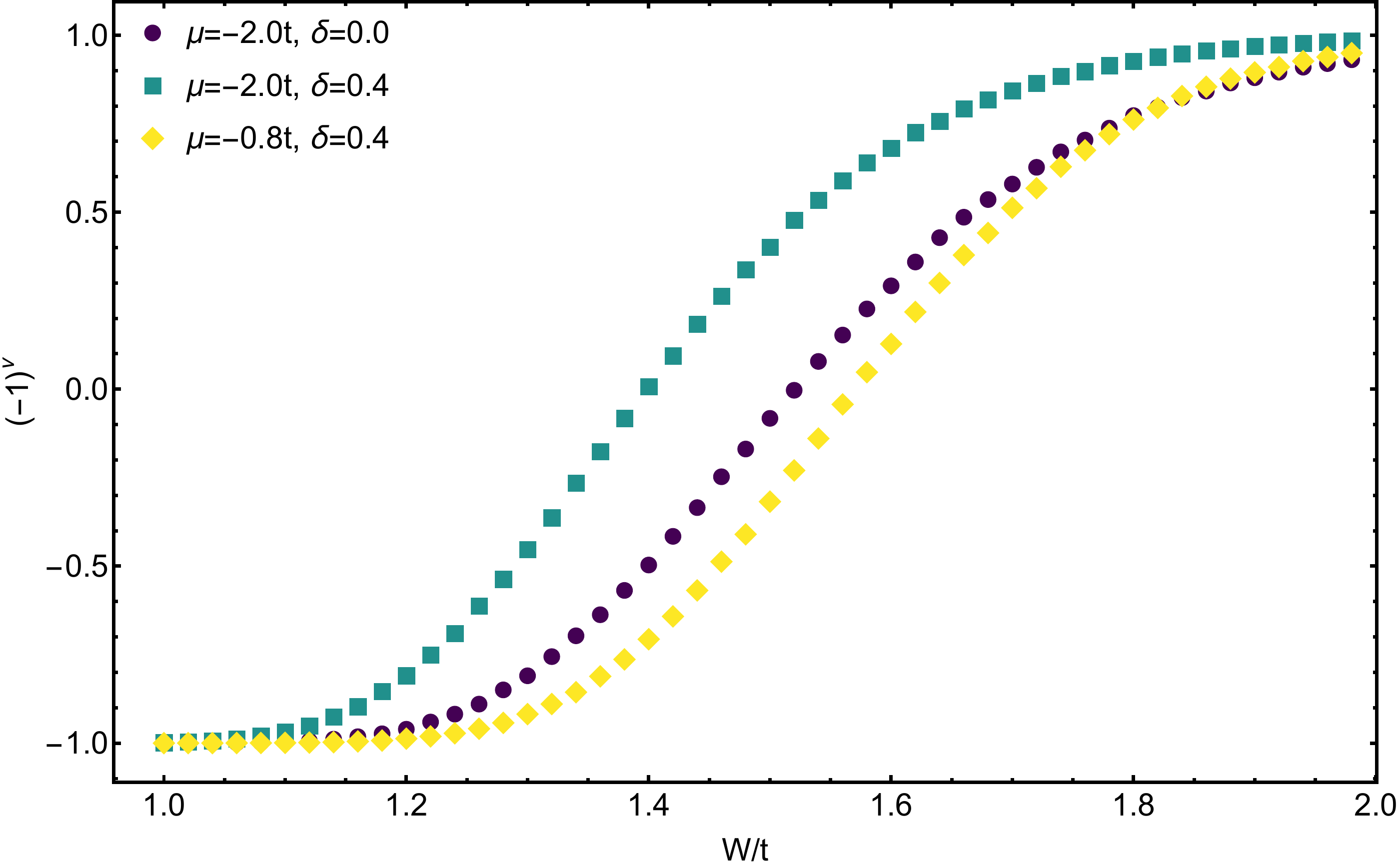}
\caption{
Topological transition driven by the electrostatic disorder for three representative values of $\mu$ and $\delta$, as indicated. Perhaps unsurprisingly introducing dimerization to the completely homogeneous case, causes the transition to occur for smaller disorder strengths. However in the dimerization-induced phase (yellow diamonds) the transition occurs at slightly larger disorder values. The rest of the parameters used in calculation were: $h=0.3t$, $\lambda = 0.15t$, and $\Delta = 0.2t$.
}
\label{Z2_vs_disorder}
\end{figure}

\begin{figure}[t]
\includegraphics[width=\linewidth,keepaspectratio]{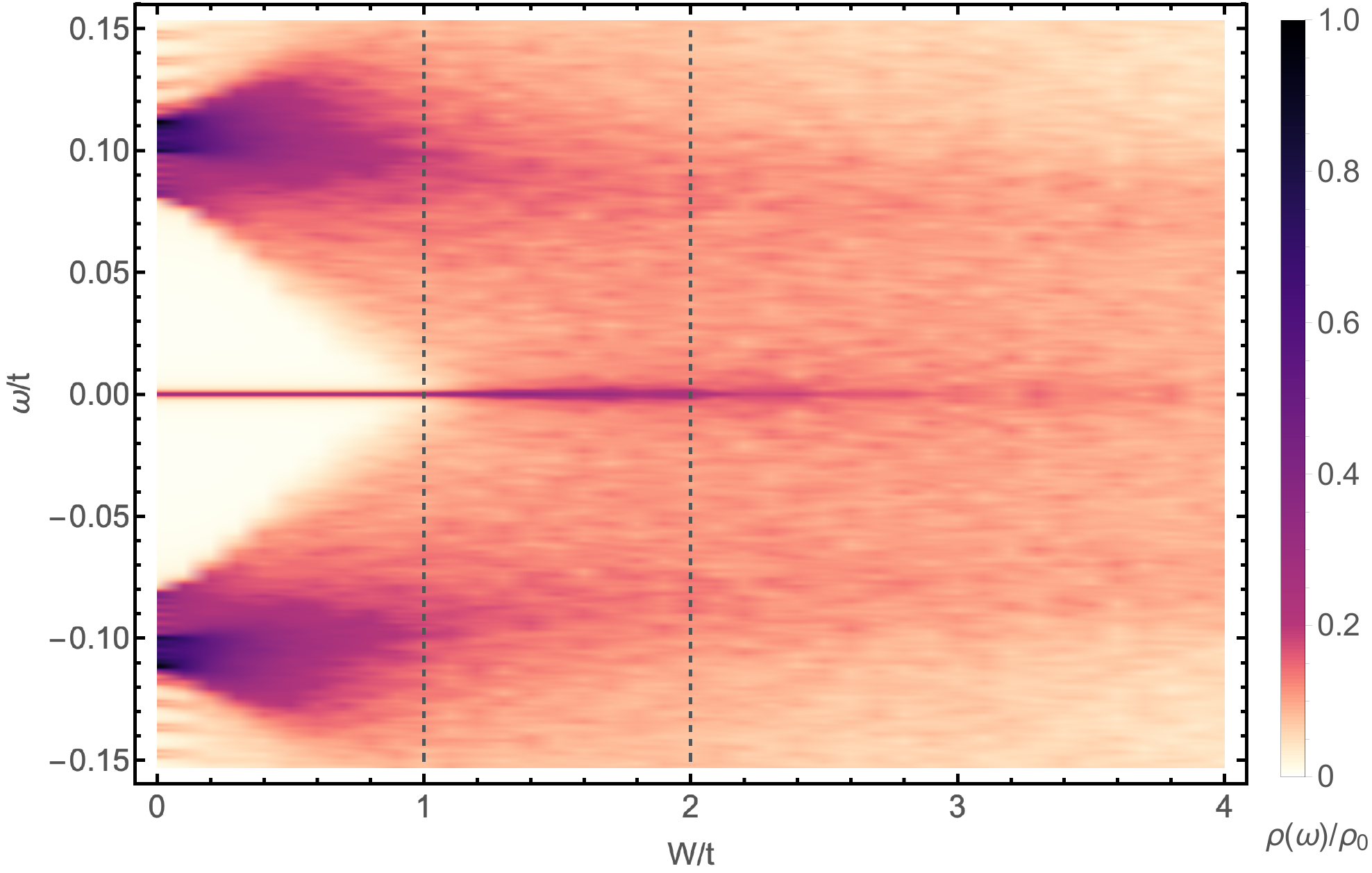}
\caption{
Changeover of the global density of states averaged over electrostatic disorder versus its amplitude $W$ obtained for $\mu=-2t$, $\delta=0.4t$, $h=0.3t$, $\lambda = 0.15t$ and $\Delta = 0.2t$. The dashed lines show the limits in which the disorder induced transition occurs on average. The MZMs can still be seen in this regime, although on average the gap has already been closed. The density of states is scaled by $\rho_0=1.26\times 10^{4}t^{-1}$.
}
\label{topoDOS_vs_disorder}
\end{figure}

Finally we check whether the topological phase driven solely by dimerization is equally stable against disorder, as the topological phase of the homogeneous nanowire. We thus introduce a random on-site term
\begin{equation}
    \mathcal{H}_{\rm dis}=\frac{W}{2}\sum_i\xi_i\Psi_i^\dagger \tau^z \Psi_i\,,
\label{random_field}
\end{equation}
where $W$ stands for the disorder amplitude and $-1\leq \xi_i\leq 1$ are random numbers. We have diagonalized the system using the Bogoliubov--de~Gennes technique and computed the quasiparticle spectra as well as the topological invariant averaged over $10^5$ different distributions $\left\{ \xi_{i}\right\}$.

Fiure.~\ref{Z2_vs_disorder} illustrates the effect of the disorder on the averaged topological invariant obtained at three representative points in the phase diagram. We have evaluated the topological index using the scattering method \cite{akhmerov.11,fulga.11} and averaged it over $10^5$ configurations of the electrostatic disorder. Although the index is constrained to be either $-1$ or $1$ for any particular disorder realization, upon averaging it shows a smooth cross-over between these values. No substantial difference in the robustness to disorder can be seen for the three cases considered by us, i.e., (i) the homogeneous non-trivial phase; (ii) a point in the continuation of this phase in the dimerized case; (iii) and a point in the dimerization induced topologically non-trivial phase. We noticed that compared to (i), case (ii) tends towards the topological crossover at smaller disorder strengths. More surprisingly in case (iii) the topological phase survives on average to larger disorder strengths. This is despite the size of the gap being slightly smaller in case (iii) than either (i) or (ii).

The change of the averaged topological index is simultaneously reflected in the local density of states (Fig.~\ref{topoDOS_vs_disorder}). The gap closes, on average, at the same disorder amplitude where the topological index begins to change its value between $-1$ and $1$. In this region, however, there still exist realizations of the random electrostatic fields (\ref{random_field}) when the MZMs survive, as is evidenced by the well pronounced spectral weight at $\omega=0$ (Fig.~\ref{topoDOS_vs_disorder}). This phenomenon partly resembles  the role played by thermal effects, as has been recently predicted for the uniform Rashba nanowires using the Monte Carlo studies \cite{maska_etal_2019}.

\section{Summary and perspectives}
\label{sec.sum}

We have studied the influence of dimerization on the topological phases of a Rashba nanowire proximitized to a superconducting substrate. We have found that sufficiently strong alternation of the hopping integral is detrimental to the topological superconducting phase, as evidenced by closing of the  protecting gap and subsequent disappearance of the Majorana zero-modes. Besides this detrimental role, however, we have also predicted an additional topological phase appearing well outside the usual regions typical for the uniform Rashba nanowires. Inspecting symmetries of the system and the related band inversion  we have analytically determined the topological invariant and constructed the phase diagrams with respect to all parameters of the model.

Our results indicate that dimerization might be beneficial for realization of the topological superconducting phase in the proximitized Rashba nanowires. In practice such a situation might be encountered, for instance, in extremely narrow metallic strips (comprising a ladder of itinerant electrons) sandwiched between two external superconducting reservoirs, analogous to what has been recently experimentally reported in Refs.~\onlinecite{Fornieri-2019,Ren-2019}. The dimerized Rashba systems could also be realized using either double- (or multi-) chain arrangements of some magnetic atoms, such as Co and Fe, weakly interconnected between themselves and deposited on surfaces of superconducting substrates~\cite{Fulga-2019}. A general approach for such tailor-made band structures could be practically achieved through atom manipulation using STM \cite{Drost-2017}.

Another feasible version of an emergent symmetry protection due to dimerization manifested in the structure of the $\mathbb{Z}_{2}$ fields can be related with the topological bond order of the interacting (correlated)  ultra cold atom systems \cite{Lewenstein-PRL2019,Lewenstein-PRB2019}.

Further theoretical studies would be useful to verify whether a tendency towards the chain dimerization is energetically favorable or unfavorable. Experimental fabrication and detection of the resulting quasiparticles in such dimerized nanosystems is also welcome and may enable a new route towards controllable manipulation of the  Majorana zero-modes.

\begin{acknowledgments}
We thank M. Lewenstein for pointing to us important role of dimerization and its influence on topological phases. This work was supported by the National Science Centre (NCN, Poland) under Grants 2018/31/N/ST3/01746 (A.K.), 2018/29/B/ST3/01892 (M.M.M.), 2017/27/B/ST3/01911 (T.D.), and 2017/27/B/ST3/02881 (N.S.). 
\end{acknowledgments}

\appendix

\section{Gap closing lines for the topological phase transitions}
\label{appendix}

The topological phase diagram of our dimerized Rashba nanowire is given by Eq.~\eqref{z2eqn_index}. Its topologically trivial and non-trivial phases must be separated by gap closing points at either $k=0$ or $k=\pi$. For $k=0$ one finds such closing at
\begin{eqnarray}
    h^2&=&4t^2-4\lambda^2\delta^2+\Delta^2+\mu^2
    \\&&\pm\:4\sqrt{t^2\mu^2-(4t^2+\Delta^2)\lambda^2\delta^2} \nonumber 
\end{eqnarray}
or, solving for the chemical potential,
\begin{eqnarray}
    \mu^2&=&4t^2+h^2+4\lambda^2\delta^2-\Delta^2
    \\ && \pm\:4\sqrt{t^2h^2-\Delta^2(t+\lambda^2\delta^2)}\,. \nonumber
\end{eqnarray}
The other closing, at $k=\pi$, occurs when
\begin{eqnarray}
    h^2&=&4t^2\delta^2-4\lambda^2+\Delta^2+\mu^2 \\ && \pm\:4\sqrt{t^2\delta^2\mu^2-\lambda^2(4t^2\delta^2+\Delta^2)} \nonumber
\end{eqnarray}
or, for the chemical potential,
\begin{eqnarray}
    \mu^2&=&h^2+4\lambda^2+4t^2\delta^2-\Delta^2 \\ && \pm\:4\sqrt{t^2h^2\delta^2-\Delta^2(\lambda^2+t^2\delta^2)}\,. \nonumber
\end{eqnarray}
In the limit of large $\mu,h>>t,\Delta,\lambda$ these expressions simplify to
\begin{equation}
    \mu^2-h^2\approx\pm\:4|th|\,,
\end{equation}
at $k=0$, and 
\begin{equation}
    \mu^2-h^2\approx\pm\:4|th\delta|\,,
\end{equation}
at $k=\pi$. The condition to be in the topologically non-trivial phase therefore becomes
\begin{equation}
  4|th\delta|\leq|\mu^2-h^2|\leq4|th|\,,
\end{equation}
in this limit.

\bibliography{biblio}

\begin{thebibliography}{83}%
\makeatletter
\providecommand \@ifxundefined [1]{%
 \@ifx{#1\undefined}
}%
\providecommand \@ifnum [1]{%
 \ifnum #1\expandafter \@firstoftwo
 \else \expandafter \@secondoftwo
 \fi
}%
\providecommand \@ifx [1]{%
 \ifx #1\expandafter \@firstoftwo
 \else \expandafter \@secondoftwo
 \fi
}%
\providecommand \natexlab [1]{#1}%
\providecommand \enquote  [1]{``#1''}%
\providecommand \bibnamefont  [1]{#1}%
\providecommand \bibfnamefont [1]{#1}%
\providecommand \citenamefont [1]{#1}%
\providecommand \href@noop [0]{\@secondoftwo}%
\providecommand \href [0]{\begingroup \@sanitize@url \@href}%
\providecommand \@href[1]{\@@startlink{#1}\@@href}%
\providecommand \@@href[1]{\endgroup#1\@@endlink}%
\providecommand \@sanitize@url [0]{\catcode `\\12\catcode `\$12\catcode
  `\&12\catcode `\#12\catcode `\^12\catcode `\_12\catcode `\%12\relax}%
\providecommand \@@startlink[1]{}%
\providecommand \@@endlink[0]{}%
\providecommand \url  [0]{\begingroup\@sanitize@url \@url }%
\providecommand \@url [1]{\endgroup\@href {#1}{\urlprefix }}%
\providecommand \urlprefix  [0]{URL }%
\providecommand \Eprint [0]{\href }%
\providecommand \doibase [0]{http://dx.doi.org/}%
\providecommand \selectlanguage [0]{\@gobble}%
\providecommand \bibinfo  [0]{\@secondoftwo}%
\providecommand \bibfield  [0]{\@secondoftwo}%
\providecommand \translation [1]{[#1]}%
\providecommand \BibitemOpen [0]{}%
\providecommand \bibitemStop [0]{}%
\providecommand \bibitemNoStop [0]{.\EOS\space}%
\providecommand \EOS [0]{\spacefactor3000\relax}%
\providecommand \BibitemShut  [1]{\csname bibitem#1\endcsname}%
\let\auto@bib@innerbib\@empty
\bibitem [{\citenamefont {Kitaev}(2001)}]{kitaev.01}%
  \BibitemOpen
  \bibfield  {author} {\bibinfo {author} {\bibfnamefont {A.~Y.}\ \bibnamefont
  {Kitaev}},\ }\href {\doibase 10.1070/1063-7869/44/10S/S29} {\bibfield
  {journal} {\bibinfo  {journal} {Phys.-Usp.}\ }\textbf {\bibinfo {volume}
  {44}},\ \bibinfo {pages} {131} (\bibinfo {year} {2001})}\BibitemShut
  {NoStop}%
\bibitem [{\citenamefont {Deng}\ \emph {et~al.}(2012)\citenamefont {Deng},
  \citenamefont {Yu}, \citenamefont {Huang}, \citenamefont {Larsson},
  \citenamefont {Caroff},\ and\ \citenamefont {Xu}}]{deng.yu.12}%
  \BibitemOpen
  \bibfield  {author} {\bibinfo {author} {\bibfnamefont {M.~T.}\ \bibnamefont
  {Deng}}, \bibinfo {author} {\bibfnamefont {C.~L.}\ \bibnamefont {Yu}},
  \bibinfo {author} {\bibfnamefont {G.~Y.}\ \bibnamefont {Huang}}, \bibinfo
  {author} {\bibfnamefont {M.}~\bibnamefont {Larsson}}, \bibinfo {author}
  {\bibfnamefont {P.}~\bibnamefont {Caroff}}, \ and\ \bibinfo {author}
  {\bibfnamefont {H.~Q.}\ \bibnamefont {Xu}},\ }\href {\doibase
  10.1021/nl303758w} {\bibfield  {journal} {\bibinfo  {journal} {Nano Lett.}\
  }\textbf {\bibinfo {volume} {12}},\ \bibinfo {pages} {6414} (\bibinfo {year}
  {2012})}\BibitemShut {NoStop}%
\bibitem [{\citenamefont {Mourik}\ \emph {et~al.}(2012)\citenamefont {Mourik},
  \citenamefont {Zuo}, \citenamefont {Frolov}, \citenamefont {Plissard},
  \citenamefont {Bakkers},\ and\ \citenamefont {Kouwenhoven}}]{mourik.zuo.12}%
  \BibitemOpen
  \bibfield  {author} {\bibinfo {author} {\bibfnamefont {V.}~\bibnamefont
  {Mourik}}, \bibinfo {author} {\bibfnamefont {K.}~\bibnamefont {Zuo}},
  \bibinfo {author} {\bibfnamefont {S.~M.}\ \bibnamefont {Frolov}}, \bibinfo
  {author} {\bibfnamefont {S.~R.}\ \bibnamefont {Plissard}}, \bibinfo {author}
  {\bibfnamefont {E.~P. A.~M.}\ \bibnamefont {Bakkers}}, \ and\ \bibinfo
  {author} {\bibfnamefont {L.~P.}\ \bibnamefont {Kouwenhoven}},\ }\href
  {\doibase 10.1126/science.1222360} {\bibfield  {journal} {\bibinfo  {journal}
  {Science}\ }\textbf {\bibinfo {volume} {336}},\ \bibinfo {pages} {1003}
  (\bibinfo {year} {2012})}\BibitemShut {NoStop}%
\bibitem [{\citenamefont {Das}\ \emph {et~al.}(2012)\citenamefont {Das},
  \citenamefont {Ronen}, \citenamefont {Most}, \citenamefont {Oreg},
  \citenamefont {Heiblum},\ and\ \citenamefont {Shtrikman}}]{das.ronen.12}%
  \BibitemOpen
  \bibfield  {author} {\bibinfo {author} {\bibfnamefont {A.}~\bibnamefont
  {Das}}, \bibinfo {author} {\bibfnamefont {Y.}~\bibnamefont {Ronen}}, \bibinfo
  {author} {\bibfnamefont {Y.}~\bibnamefont {Most}}, \bibinfo {author}
  {\bibfnamefont {Y.}~\bibnamefont {Oreg}}, \bibinfo {author} {\bibfnamefont
  {M.}~\bibnamefont {Heiblum}}, \ and\ \bibinfo {author} {\bibfnamefont
  {H.}~\bibnamefont {Shtrikman}},\ }\href {\doibase 10.1038/nphys2479}
  {\bibfield  {journal} {\bibinfo  {journal} {Nat. Phys.}\ }\textbf {\bibinfo
  {volume} {8}},\ \bibinfo {pages} {887} (\bibinfo {year} {2012})}\BibitemShut
  {NoStop}%
\bibitem [{\citenamefont {Finck}\ \emph {et~al.}(2013)\citenamefont {Finck},
  \citenamefont {Van~Harlingen}, \citenamefont {Mohseni}, \citenamefont
  {Jung},\ and\ \citenamefont {Li}}]{finck.vanharlingen.13}%
  \BibitemOpen
  \bibfield  {author} {\bibinfo {author} {\bibfnamefont {A.~D.~K.}\
  \bibnamefont {Finck}}, \bibinfo {author} {\bibfnamefont {D.~J.}\ \bibnamefont
  {Van~Harlingen}}, \bibinfo {author} {\bibfnamefont {P.~K.}\ \bibnamefont
  {Mohseni}}, \bibinfo {author} {\bibfnamefont {K.}~\bibnamefont {Jung}}, \
  and\ \bibinfo {author} {\bibfnamefont {X.}~\bibnamefont {Li}},\ }\href
  {\doibase 10.1103/PhysRevLett.110.126406} {\bibfield  {journal} {\bibinfo
  {journal} {Phys. Rev. Lett.}\ }\textbf {\bibinfo {volume} {110}},\ \bibinfo
  {pages} {126406} (\bibinfo {year} {2013})}\BibitemShut {NoStop}%
\bibitem [{\citenamefont {Deng}\ \emph {et~al.}(2016)\citenamefont {Deng},
  \citenamefont {Vaitiekenas}, \citenamefont {Hansen}, \citenamefont {Danon},
  \citenamefont {Leijnse}, \citenamefont {Flensberg}, \citenamefont {Nyg{\r
  a}rd}, \citenamefont {Krogstrup},\ and\ \citenamefont
  {Marcus}}]{deng.vaitiekenas.16}%
  \BibitemOpen
  \bibfield  {author} {\bibinfo {author} {\bibfnamefont {M.~T.}\ \bibnamefont
  {Deng}}, \bibinfo {author} {\bibfnamefont {S.}~\bibnamefont {Vaitiekenas}},
  \bibinfo {author} {\bibfnamefont {E.~B.}\ \bibnamefont {Hansen}}, \bibinfo
  {author} {\bibfnamefont {J.}~\bibnamefont {Danon}}, \bibinfo {author}
  {\bibfnamefont {M.}~\bibnamefont {Leijnse}}, \bibinfo {author} {\bibfnamefont
  {K.}~\bibnamefont {Flensberg}}, \bibinfo {author} {\bibfnamefont
  {J.}~\bibnamefont {Nyg{\r a}rd}}, \bibinfo {author} {\bibfnamefont
  {P.}~\bibnamefont {Krogstrup}}, \ and\ \bibinfo {author} {\bibfnamefont
  {C.~M.}\ \bibnamefont {Marcus}},\ }\href {\doibase 10.1126/science.aaf3961}
  {\bibfield  {journal} {\bibinfo  {journal} {Science}\ }\textbf {\bibinfo
  {volume} {354}},\ \bibinfo {pages} {1557} (\bibinfo {year}
  {2016})}\BibitemShut {NoStop}%
\bibitem [{\citenamefont {Sestoft}\ \emph {et~al.}(2018)\citenamefont
  {Sestoft}, \citenamefont {Kanne}, \citenamefont {Gejl}, \citenamefont {von
  Soosten}, \citenamefont {Yodh}, \citenamefont {Sherman}, \citenamefont
  {Tarasinski}, \citenamefont {Wimmer}, \citenamefont {Johnson}, \citenamefont
  {Deng}, \citenamefont {Nygard}, \citenamefont {Jespersen}, \citenamefont
  {Marcus},\ and\ \citenamefont {Krogstrup}}]{seastoft.kanne.18}%
  \BibitemOpen
  \bibfield  {author} {\bibinfo {author} {\bibfnamefont {J.~E.}\ \bibnamefont
  {Sestoft}}, \bibinfo {author} {\bibfnamefont {T.}~\bibnamefont {Kanne}},
  \bibinfo {author} {\bibfnamefont {A.~N.}\ \bibnamefont {Gejl}}, \bibinfo
  {author} {\bibfnamefont {M.}~\bibnamefont {von Soosten}}, \bibinfo {author}
  {\bibfnamefont {J.~S.}\ \bibnamefont {Yodh}}, \bibinfo {author}
  {\bibfnamefont {D.}~\bibnamefont {Sherman}}, \bibinfo {author} {\bibfnamefont
  {B.}~\bibnamefont {Tarasinski}}, \bibinfo {author} {\bibfnamefont
  {M.}~\bibnamefont {Wimmer}}, \bibinfo {author} {\bibfnamefont
  {E.}~\bibnamefont {Johnson}}, \bibinfo {author} {\bibfnamefont
  {M.}~\bibnamefont {Deng}}, \bibinfo {author} {\bibfnamefont {J.}~\bibnamefont
  {Nygard}}, \bibinfo {author} {\bibfnamefont {T.~S.}\ \bibnamefont
  {Jespersen}}, \bibinfo {author} {\bibfnamefont {C.~M.}\ \bibnamefont
  {Marcus}}, \ and\ \bibinfo {author} {\bibfnamefont {P.}~\bibnamefont
  {Krogstrup}},\ }\href {\doibase 10.1103/PhysRevMaterials.2.044202} {\bibfield
   {journal} {\bibinfo  {journal} {Phys. Rev. Mater.}\ }\textbf {\bibinfo
  {volume} {2}},\ \bibinfo {pages} {44202} (\bibinfo {year}
  {2018})}\BibitemShut {NoStop}%
\bibitem [{\citenamefont {Lutchyn}\ \emph {et~al.}(2018)\citenamefont
  {Lutchyn}, \citenamefont {Bakkers}, \citenamefont {Kouwenhoven},
  \citenamefont {Krogstrup}, \citenamefont {Marcus},\ and\ \citenamefont
  {Oreg}}]{lutchyn.bakkers.18}%
  \BibitemOpen
  \bibfield  {author} {\bibinfo {author} {\bibfnamefont {R.~M.}\ \bibnamefont
  {Lutchyn}}, \bibinfo {author} {\bibfnamefont {E.~P. A.~M.}\ \bibnamefont
  {Bakkers}}, \bibinfo {author} {\bibfnamefont {L.~P.}\ \bibnamefont
  {Kouwenhoven}}, \bibinfo {author} {\bibfnamefont {P.}~\bibnamefont
  {Krogstrup}}, \bibinfo {author} {\bibfnamefont {C.~M.}\ \bibnamefont
  {Marcus}}, \ and\ \bibinfo {author} {\bibfnamefont {Y.}~\bibnamefont
  {Oreg}},\ }\href {\doibase 10.1038/s41578-018-0003-1} {\bibfield  {journal}
  {\bibinfo  {journal} {Nat. Rev. Mater.}\ }\textbf {\bibinfo {volume} {3}},\
  \bibinfo {pages} {52} (\bibinfo {year} {2018})}\BibitemShut {NoStop}%
\bibitem [{\citenamefont {G\"ul}\ \emph {et~al.}(2018)\citenamefont {G\"ul},
  \citenamefont {Zhang}, \citenamefont {Bommer}, \citenamefont {de~Moor},
  \citenamefont {Car}, \citenamefont {Plissard}, \citenamefont {Bakkers},
  \citenamefont {Geresdi}, \citenamefont {Watanabe}, \citenamefont
  {Taniguchi},\ and\ \citenamefont {Kouwenhoven}}]{gul.zhang.18}%
  \BibitemOpen
  \bibfield  {author} {\bibinfo {author} {\bibfnamefont {O.}~\bibnamefont
  {G\"ul}}, \bibinfo {author} {\bibfnamefont {H.}~\bibnamefont {Zhang}},
  \bibinfo {author} {\bibfnamefont {J.}~\bibnamefont {Bommer}}, \bibinfo
  {author} {\bibfnamefont {M.}~\bibnamefont {de~Moor}}, \bibinfo {author}
  {\bibfnamefont {D.}~\bibnamefont {Car}}, \bibinfo {author} {\bibfnamefont
  {S.}~\bibnamefont {Plissard}}, \bibinfo {author} {\bibfnamefont
  {E.}~\bibnamefont {Bakkers}}, \bibinfo {author} {\bibfnamefont
  {A.}~\bibnamefont {Geresdi}}, \bibinfo {author} {\bibfnamefont
  {K.}~\bibnamefont {Watanabe}}, \bibinfo {author} {\bibfnamefont
  {T.}~\bibnamefont {Taniguchi}}, \ and\ \bibinfo {author} {\bibfnamefont
  {L.}~\bibnamefont {Kouwenhoven}},\ }\href {\doibase
  10.1038/s41565-017-0032-8} {\bibfield  {journal} {\bibinfo  {journal} {Nat.
  Nanotechnol.}\ }\textbf {\bibinfo {volume} {13}},\ \bibinfo {pages} {192}
  (\bibinfo {year} {2018})}\BibitemShut {NoStop}%
\bibitem [{\citenamefont {Nadj-Perge}\ \emph {et~al.}(2014)\citenamefont
  {Nadj-Perge}, \citenamefont {Drozdov}, \citenamefont {Li}, \citenamefont
  {Chen}, \citenamefont {Jeon}, \citenamefont {Seo}, \citenamefont {MacDonald},
  \citenamefont {Bernevig},\ and\ \citenamefont
  {Yazdani}}]{nadjperge.drozdov.14}%
  \BibitemOpen
  \bibfield  {author} {\bibinfo {author} {\bibfnamefont {S.}~\bibnamefont
  {Nadj-Perge}}, \bibinfo {author} {\bibfnamefont {I.~K.}\ \bibnamefont
  {Drozdov}}, \bibinfo {author} {\bibfnamefont {J.}~\bibnamefont {Li}},
  \bibinfo {author} {\bibfnamefont {H.}~\bibnamefont {Chen}}, \bibinfo {author}
  {\bibfnamefont {S.}~\bibnamefont {Jeon}}, \bibinfo {author} {\bibfnamefont
  {J.}~\bibnamefont {Seo}}, \bibinfo {author} {\bibfnamefont {A.~H.}\
  \bibnamefont {MacDonald}}, \bibinfo {author} {\bibfnamefont {B.~A.}\
  \bibnamefont {Bernevig}}, \ and\ \bibinfo {author} {\bibfnamefont
  {A.}~\bibnamefont {Yazdani}},\ }\href {\doibase 10.1126/science.1259327}
  {\bibfield  {journal} {\bibinfo  {journal} {Science}\ }\textbf {\bibinfo
  {volume} {346}},\ \bibinfo {pages} {602} (\bibinfo {year}
  {2014})}\BibitemShut {NoStop}%
\bibitem [{\citenamefont {Pawlak}\ \emph {et~al.}(2016)\citenamefont {Pawlak},
  \citenamefont {Kisiel}, \citenamefont {Klinovaja}, \citenamefont {Meier},
  \citenamefont {Kawai}, \citenamefont {Glatzel}, \citenamefont {Loss},\ and\
  \citenamefont {Meyer}}]{pawlak.kisiel.16}%
  \BibitemOpen
  \bibfield  {author} {\bibinfo {author} {\bibfnamefont {R.}~\bibnamefont
  {Pawlak}}, \bibinfo {author} {\bibfnamefont {M.}~\bibnamefont {Kisiel}},
  \bibinfo {author} {\bibfnamefont {J.}~\bibnamefont {Klinovaja}}, \bibinfo
  {author} {\bibfnamefont {T.}~\bibnamefont {Meier}}, \bibinfo {author}
  {\bibfnamefont {S.}~\bibnamefont {Kawai}}, \bibinfo {author} {\bibfnamefont
  {T.}~\bibnamefont {Glatzel}}, \bibinfo {author} {\bibfnamefont
  {D.}~\bibnamefont {Loss}}, \ and\ \bibinfo {author} {\bibfnamefont
  {E.}~\bibnamefont {Meyer}},\ }\href {\doibase 10.1038/npjqi.2016.35}
  {\bibfield  {journal} {\bibinfo  {journal} {Npj Quantum Information}\
  }\textbf {\bibinfo {volume} {2}},\ \bibinfo {pages} {16035} (\bibinfo {year}
  {2016})}\BibitemShut {NoStop}%
\bibitem [{\citenamefont {Feldman}\ \emph {et~al.}(2016)\citenamefont
  {Feldman}, \citenamefont {Randeria}, \citenamefont {Li}, \citenamefont
  {Jeon}, \citenamefont {Xie}, \citenamefont {Wang}, \citenamefont {Drozdov},
  \citenamefont {Bernevig},\ and\ \citenamefont
  {Yazdani}}]{feldman.randeria.16}%
  \BibitemOpen
  \bibfield  {author} {\bibinfo {author} {\bibfnamefont {B.~E.}\ \bibnamefont
  {Feldman}}, \bibinfo {author} {\bibfnamefont {M.~T.}\ \bibnamefont
  {Randeria}}, \bibinfo {author} {\bibfnamefont {J.}~\bibnamefont {Li}},
  \bibinfo {author} {\bibfnamefont {S.}~\bibnamefont {Jeon}}, \bibinfo {author}
  {\bibfnamefont {Y.}~\bibnamefont {Xie}}, \bibinfo {author} {\bibfnamefont
  {Z.}~\bibnamefont {Wang}}, \bibinfo {author} {\bibfnamefont {I.~K.}\
  \bibnamefont {Drozdov}}, \bibinfo {author} {\bibfnamefont {B.~A.}\
  \bibnamefont {Bernevig}}, \ and\ \bibinfo {author} {\bibfnamefont
  {A.}~\bibnamefont {Yazdani}},\ }\href {\doibase 10.1038/nphys3947} {\bibfield
   {journal} {\bibinfo  {journal} {Nat. Phys.}\ }\textbf {\bibinfo {volume}
  {13}},\ \bibinfo {pages} {286} (\bibinfo {year} {2016})}\BibitemShut
  {NoStop}%
\bibitem [{\citenamefont {Ruby}\ \emph {et~al.}(2017)\citenamefont {Ruby},
  \citenamefont {Heinrich}, \citenamefont {Peng}, \citenamefont {von Oppen},\
  and\ \citenamefont {Franke}}]{ruby.heinrich.17}%
  \BibitemOpen
  \bibfield  {author} {\bibinfo {author} {\bibfnamefont {M.}~\bibnamefont
  {Ruby}}, \bibinfo {author} {\bibfnamefont {B.~W.}\ \bibnamefont {Heinrich}},
  \bibinfo {author} {\bibfnamefont {Y.}~\bibnamefont {Peng}}, \bibinfo {author}
  {\bibfnamefont {F.}~\bibnamefont {von Oppen}}, \ and\ \bibinfo {author}
  {\bibfnamefont {K.~J.}\ \bibnamefont {Franke}},\ }\href {\doibase
  10.1021/acs.nanolett.7b01728} {\bibfield  {journal} {\bibinfo  {journal}
  {Nano Lett.}\ }\textbf {\bibinfo {volume} {17}},\ \bibinfo {pages} {4473}
  (\bibinfo {year} {2017})}\BibitemShut {NoStop}%
\bibitem [{\citenamefont {Jeon}\ \emph {et~al.}(2017)\citenamefont {Jeon},
  \citenamefont {Xie}, \citenamefont {Li}, \citenamefont {Wang}, \citenamefont
  {Bernevig},\ and\ \citenamefont {Yazdani}}]{jeon.xie.17}%
  \BibitemOpen
  \bibfield  {author} {\bibinfo {author} {\bibfnamefont {S.}~\bibnamefont
  {Jeon}}, \bibinfo {author} {\bibfnamefont {Y.}~\bibnamefont {Xie}}, \bibinfo
  {author} {\bibfnamefont {J.}~\bibnamefont {Li}}, \bibinfo {author}
  {\bibfnamefont {Z.}~\bibnamefont {Wang}}, \bibinfo {author} {\bibfnamefont
  {B.~A.}\ \bibnamefont {Bernevig}}, \ and\ \bibinfo {author} {\bibfnamefont
  {A.}~\bibnamefont {Yazdani}},\ }\href {\doibase 10.1126/science.aan3670}
  {\bibfield  {journal} {\bibinfo  {journal} {Science}\ }\textbf {\bibinfo
  {volume} {358}},\ \bibinfo {pages} {772} (\bibinfo {year}
  {2017})}\BibitemShut {NoStop}%
\bibitem [{\citenamefont {Kim}\ \emph {et~al.}(2018)\citenamefont {Kim},
  \citenamefont {Palacio-Morales}, \citenamefont {Posske}, \citenamefont
  {R{\'o}zsa}, \citenamefont {Palot{\'a}s}, \citenamefont {Szunyogh},
  \citenamefont {Thorwart},\ and\ \citenamefont
  {Wiesendanger}}]{kim.palaciomorales.18}%
  \BibitemOpen
  \bibfield  {author} {\bibinfo {author} {\bibfnamefont {H.}~\bibnamefont
  {Kim}}, \bibinfo {author} {\bibfnamefont {A.}~\bibnamefont
  {Palacio-Morales}}, \bibinfo {author} {\bibfnamefont {T.}~\bibnamefont
  {Posske}}, \bibinfo {author} {\bibfnamefont {L.}~\bibnamefont {R{\'o}zsa}},
  \bibinfo {author} {\bibfnamefont {K.}~\bibnamefont {Palot{\'a}s}}, \bibinfo
  {author} {\bibfnamefont {L.}~\bibnamefont {Szunyogh}}, \bibinfo {author}
  {\bibfnamefont {M.}~\bibnamefont {Thorwart}}, \ and\ \bibinfo {author}
  {\bibfnamefont {R.}~\bibnamefont {Wiesendanger}},\ }\href
  {https://advances.sciencemag.org/content/4/5/eaar5251} {\bibfield  {journal}
  {\bibinfo  {journal} {Sci. Adv.}\ }\textbf {\bibinfo {volume} {4}},\ \bibinfo
  {pages} {eaar5251} (\bibinfo {year} {2018})}\BibitemShut {NoStop}%
\bibitem [{\citenamefont {Nichele}\ \emph {et~al.}(2017)\citenamefont
  {Nichele}, \citenamefont {Drachmann}, \citenamefont {Whiticar}, \citenamefont
  {O'Farrell}, \citenamefont {Suominen}, \citenamefont {Fornieri},
  \citenamefont {Wang}, \citenamefont {Gardner}, \citenamefont {Thomas},
  \citenamefont {Hatke}, \citenamefont {Krogstrup}, \citenamefont {Manfra},
  \citenamefont {Flensberg},\ and\ \citenamefont
  {Marcus}}]{nichele.ofarrell.17}%
  \BibitemOpen
  \bibfield  {author} {\bibinfo {author} {\bibfnamefont {F.}~\bibnamefont
  {Nichele}}, \bibinfo {author} {\bibfnamefont {A.~C.~C.}\ \bibnamefont
  {Drachmann}}, \bibinfo {author} {\bibfnamefont {A.~M.}\ \bibnamefont
  {Whiticar}}, \bibinfo {author} {\bibfnamefont {E.~C.~T.}\ \bibnamefont
  {O'Farrell}}, \bibinfo {author} {\bibfnamefont {H.~J.}\ \bibnamefont
  {Suominen}}, \bibinfo {author} {\bibfnamefont {A.}~\bibnamefont {Fornieri}},
  \bibinfo {author} {\bibfnamefont {T.}~\bibnamefont {Wang}}, \bibinfo {author}
  {\bibfnamefont {G.~C.}\ \bibnamefont {Gardner}}, \bibinfo {author}
  {\bibfnamefont {C.}~\bibnamefont {Thomas}}, \bibinfo {author} {\bibfnamefont
  {A.~T.}\ \bibnamefont {Hatke}}, \bibinfo {author} {\bibfnamefont
  {P.}~\bibnamefont {Krogstrup}}, \bibinfo {author} {\bibfnamefont {M.~J.}\
  \bibnamefont {Manfra}}, \bibinfo {author} {\bibfnamefont {K.}~\bibnamefont
  {Flensberg}}, \ and\ \bibinfo {author} {\bibfnamefont {C.~M.}\ \bibnamefont
  {Marcus}},\ }\href {\doibase 10.1103/PhysRevLett.119.136803} {\bibfield
  {journal} {\bibinfo  {journal} {Phys. Rev. Lett.}\ }\textbf {\bibinfo
  {volume} {119}},\ \bibinfo {pages} {136803} (\bibinfo {year}
  {2017})}\BibitemShut {NoStop}%
\bibitem [{\citenamefont {Fornieri}\ \emph {et~al.}(2019)\citenamefont
  {Fornieri}, \citenamefont {Whiticar}, \citenamefont {Setiawan}, \citenamefont
  {Portoles}, \citenamefont {Drachmann}, \citenamefont {Keselman},
  \citenamefont {Gronin}, \citenamefont {Thomas}, \citenamefont {Wang},
  \citenamefont {Kallaher}, \citenamefont {Gardner}, \citenamefont {Berg},
  \citenamefont {Manfra}, \citenamefont {Stern}, \citenamefont {Marcus},\ and\
  \citenamefont {Nichele}}]{Fornieri-2019}%
  \BibitemOpen
  \bibfield  {author} {\bibinfo {author} {\bibfnamefont {A.}~\bibnamefont
  {Fornieri}}, \bibinfo {author} {\bibfnamefont {A.}~\bibnamefont {Whiticar}},
  \bibinfo {author} {\bibfnamefont {F.}~\bibnamefont {Setiawan}}, \bibinfo
  {author} {\bibfnamefont {E.}~\bibnamefont {Portoles}}, \bibinfo {author}
  {\bibfnamefont {A.}~\bibnamefont {Drachmann}}, \bibinfo {author}
  {\bibfnamefont {A.}~\bibnamefont {Keselman}}, \bibinfo {author}
  {\bibfnamefont {S.}~\bibnamefont {Gronin}}, \bibinfo {author} {\bibfnamefont
  {C.}~\bibnamefont {Thomas}}, \bibinfo {author} {\bibfnamefont
  {T.}~\bibnamefont {Wang}}, \bibinfo {author} {\bibfnamefont {R.}~\bibnamefont
  {Kallaher}}, \bibinfo {author} {\bibfnamefont {G.}~\bibnamefont {Gardner}},
  \bibinfo {author} {\bibfnamefont {E.}~\bibnamefont {Berg}}, \bibinfo {author}
  {\bibfnamefont {M.}~\bibnamefont {Manfra}}, \bibinfo {author} {\bibfnamefont
  {A.}~\bibnamefont {Stern}}, \bibinfo {author} {\bibfnamefont
  {C.}~\bibnamefont {Marcus}}, \ and\ \bibinfo {author} {\bibfnamefont
  {F.}~\bibnamefont {Nichele}},\ }\href {\doibase 10.1038/s41586-019-1068-8}
  {\bibfield  {journal} {\bibinfo  {journal} {Nature}\ }\textbf {\bibinfo
  {volume} {569}},\ \bibinfo {pages} {89} (\bibinfo {year} {2019})}\BibitemShut
  {NoStop}%
\bibitem [{\citenamefont {Ren}\ \emph {et~al.}(2019)\citenamefont {Ren},
  \citenamefont {Pientka}, \citenamefont {Hart}, \citenamefont {Pierce},
  \citenamefont {Kosowsky}, \citenamefont {Lunczer}, \citenamefont {Schlereth},
  \citenamefont {Scharf}, \citenamefont {Hankiewicz}, \citenamefont
  {Molenkamp}, \citenamefont {Halperin},\ and\ \citenamefont
  {Yacoby}}]{Ren-2019}%
  \BibitemOpen
  \bibfield  {author} {\bibinfo {author} {\bibfnamefont {H.}~\bibnamefont
  {Ren}}, \bibinfo {author} {\bibfnamefont {F.}~\bibnamefont {Pientka}},
  \bibinfo {author} {\bibfnamefont {S.}~\bibnamefont {Hart}}, \bibinfo {author}
  {\bibfnamefont {A.}~\bibnamefont {Pierce}}, \bibinfo {author} {\bibfnamefont
  {M.}~\bibnamefont {Kosowsky}}, \bibinfo {author} {\bibfnamefont
  {L.}~\bibnamefont {Lunczer}}, \bibinfo {author} {\bibfnamefont
  {R.}~\bibnamefont {Schlereth}}, \bibinfo {author} {\bibfnamefont
  {B.}~\bibnamefont {Scharf}}, \bibinfo {author} {\bibfnamefont
  {E.}~\bibnamefont {Hankiewicz}}, \bibinfo {author} {\bibfnamefont
  {L.}~\bibnamefont {Molenkamp}}, \bibinfo {author} {\bibfnamefont
  {B.}~\bibnamefont {Halperin}}, \ and\ \bibinfo {author} {\bibfnamefont
  {A.}~\bibnamefont {Yacoby}},\ }\href {\doibase 10.1038/s41586-019-1148-9}
  {\bibfield  {journal} {\bibinfo  {journal} {Nature}\ }\textbf {\bibinfo
  {volume} {569}},\ \bibinfo {pages} {93} (\bibinfo {year} {2019})}\BibitemShut
  {NoStop}%
\bibitem [{\citenamefont {Sato}\ and\ \citenamefont
  {Fujimoto}(2009)}]{sato.fujimoto.09}%
  \BibitemOpen
  \bibfield  {author} {\bibinfo {author} {\bibfnamefont {M.}~\bibnamefont
  {Sato}}\ and\ \bibinfo {author} {\bibfnamefont {S.}~\bibnamefont
  {Fujimoto}},\ }\href {\doibase 10.1103/PhysRevB.79.094504} {\bibfield
  {journal} {\bibinfo  {journal} {Phys. Rev. B}\ }\textbf {\bibinfo {volume}
  {79}},\ \bibinfo {pages} {094504} (\bibinfo {year} {2009})}\BibitemShut
  {NoStop}%
\bibitem [{\citenamefont {Sato}\ \emph {et~al.}(2009)\citenamefont {Sato},
  \citenamefont {Takahashi},\ and\ \citenamefont
  {Fujimoto}}]{sato.takahashi.09}%
  \BibitemOpen
  \bibfield  {author} {\bibinfo {author} {\bibfnamefont {M.}~\bibnamefont
  {Sato}}, \bibinfo {author} {\bibfnamefont {Y.}~\bibnamefont {Takahashi}}, \
  and\ \bibinfo {author} {\bibfnamefont {S.}~\bibnamefont {Fujimoto}},\ }\href
  {\doibase 10.1103/PhysRevLett.103.020401} {\bibfield  {journal} {\bibinfo
  {journal} {Phys. Rev. Lett.}\ }\textbf {\bibinfo {volume} {103}},\ \bibinfo
  {pages} {020401} (\bibinfo {year} {2009})}\BibitemShut {NoStop}%
\bibitem [{\citenamefont {Sato}\ \emph {et~al.}(2010)\citenamefont {Sato},
  \citenamefont {Takahashi},\ and\ \citenamefont
  {Fujimoto}}]{sato.takahashi.10}%
  \BibitemOpen
  \bibfield  {author} {\bibinfo {author} {\bibfnamefont {M.}~\bibnamefont
  {Sato}}, \bibinfo {author} {\bibfnamefont {Y.}~\bibnamefont {Takahashi}}, \
  and\ \bibinfo {author} {\bibfnamefont {S.}~\bibnamefont {Fujimoto}},\ }\href
  {\doibase 10.1103/PhysRevB.82.134521} {\bibfield  {journal} {\bibinfo
  {journal} {Phys. Rev. B}\ }\textbf {\bibinfo {volume} {82}},\ \bibinfo
  {pages} {134521} (\bibinfo {year} {2010})}\BibitemShut {NoStop}%
\bibitem [{\citenamefont {Lutchyn}\ \emph {et~al.}(2010)\citenamefont
  {Lutchyn}, \citenamefont {Sau},\ and\ \citenamefont
  {Das~Sarma}}]{Lutchyn2010}%
  \BibitemOpen
  \bibfield  {author} {\bibinfo {author} {\bibfnamefont {R.~M.}\ \bibnamefont
  {Lutchyn}}, \bibinfo {author} {\bibfnamefont {J.~D.}\ \bibnamefont {Sau}}, \
  and\ \bibinfo {author} {\bibfnamefont {S.}~\bibnamefont {Das~Sarma}},\ }\href
  {\doibase 10.1103/PhysRevLett.105.077001} {\bibfield  {journal} {\bibinfo
  {journal} {Phys. Rev. Lett.}\ }\textbf {\bibinfo {volume} {105}},\ \bibinfo
  {pages} {77001} (\bibinfo {year} {2010})}\BibitemShut {NoStop}%
\bibitem [{\citenamefont {Oreg}\ \emph {et~al.}(2010)\citenamefont {Oreg},
  \citenamefont {Refael},\ and\ \citenamefont {von Oppen}}]{oreg.refael.10}%
  \BibitemOpen
  \bibfield  {author} {\bibinfo {author} {\bibfnamefont {Y.}~\bibnamefont
  {Oreg}}, \bibinfo {author} {\bibfnamefont {G.}~\bibnamefont {Refael}}, \ and\
  \bibinfo {author} {\bibfnamefont {F.}~\bibnamefont {von Oppen}},\ }\href
  {\doibase 10.1103/PhysRevLett.105.177002} {\bibfield  {journal} {\bibinfo
  {journal} {Phys. Rev. Lett.}\ }\textbf {\bibinfo {volume} {105}},\ \bibinfo
  {pages} {177002} (\bibinfo {year} {2010})}\BibitemShut {NoStop}%
\bibitem [{\citenamefont {Choy}\ \emph {et~al.}(2011)\citenamefont {Choy},
  \citenamefont {Edge}, \citenamefont {Akhmerov},\ and\ \citenamefont
  {Beenakker}}]{choy.11}%
  \BibitemOpen
  \bibfield  {author} {\bibinfo {author} {\bibfnamefont {T.-P.}\ \bibnamefont
  {Choy}}, \bibinfo {author} {\bibfnamefont {J.~M.}\ \bibnamefont {Edge}},
  \bibinfo {author} {\bibfnamefont {A.~R.}\ \bibnamefont {Akhmerov}}, \ and\
  \bibinfo {author} {\bibfnamefont {C.~W.~J.}\ \bibnamefont {Beenakker}},\
  }\href {\doibase 10.1103/PhysRevB.84.195442} {\bibfield  {journal} {\bibinfo
  {journal} {Phys. Rev. B}\ }\textbf {\bibinfo {volume} {84}},\ \bibinfo
  {pages} {195442} (\bibinfo {year} {2011})}\BibitemShut {NoStop}%
\bibitem [{\citenamefont {Martin}\ and\ \citenamefont
  {Morpurgo}(2012)}]{Martin.Morpurgo.12}%
  \BibitemOpen
  \bibfield  {author} {\bibinfo {author} {\bibfnamefont {I.}~\bibnamefont
  {Martin}}\ and\ \bibinfo {author} {\bibfnamefont {A.~F.}\ \bibnamefont
  {Morpurgo}},\ }\href {\doibase 10.1103/PhysRevB.85.144505} {\bibfield
  {journal} {\bibinfo  {journal} {Phys. Rev. B}\ }\textbf {\bibinfo {volume}
  {85}},\ \bibinfo {pages} {144505} (\bibinfo {year} {2012})}\BibitemShut
  {NoStop}%
\bibitem [{\citenamefont {Kjaergaard}\ \emph {et~al.}(2012)\citenamefont
  {Kjaergaard}, \citenamefont {W\"olms},\ and\ \citenamefont
  {Flensberg}}]{Kjaergaard2012}%
  \BibitemOpen
  \bibfield  {author} {\bibinfo {author} {\bibfnamefont {M.}~\bibnamefont
  {Kjaergaard}}, \bibinfo {author} {\bibfnamefont {K.}~\bibnamefont {W\"olms}},
  \ and\ \bibinfo {author} {\bibfnamefont {K.}~\bibnamefont {Flensberg}},\
  }\href {\doibase 10.1103/PhysRevB.85.020503} {\bibfield  {journal} {\bibinfo
  {journal} {Phys. Rev. B}\ }\textbf {\bibinfo {volume} {85}},\ \bibinfo
  {pages} {020503} (\bibinfo {year} {2012})}\BibitemShut {NoStop}%
\bibitem [{\citenamefont {Nadj-Perge}\ \emph
  {et~al.}(2013{\natexlab{a}})\citenamefont {Nadj-Perge}, \citenamefont
  {Drozdov}, \citenamefont {Bernevig},\ and\ \citenamefont
  {Yazdani}}]{Bernevig2013}%
  \BibitemOpen
  \bibfield  {author} {\bibinfo {author} {\bibfnamefont {S.}~\bibnamefont
  {Nadj-Perge}}, \bibinfo {author} {\bibfnamefont {I.~K.}\ \bibnamefont
  {Drozdov}}, \bibinfo {author} {\bibfnamefont {B.~A.}\ \bibnamefont
  {Bernevig}}, \ and\ \bibinfo {author} {\bibfnamefont {A.}~\bibnamefont
  {Yazdani}},\ }\href {\doibase 10.1103/PhysRevB.88.020407} {\bibfield
  {journal} {\bibinfo  {journal} {Phys. Rev. B}\ }\textbf {\bibinfo {volume}
  {88}},\ \bibinfo {pages} {020407} (\bibinfo {year}
  {2013}{\natexlab{a}})}\BibitemShut {NoStop}%
\bibitem [{\citenamefont {Braunecker}\ and\ \citenamefont
  {Simon}(2013)}]{Simon2013}%
  \BibitemOpen
  \bibfield  {author} {\bibinfo {author} {\bibfnamefont {B.}~\bibnamefont
  {Braunecker}}\ and\ \bibinfo {author} {\bibfnamefont {P.}~\bibnamefont
  {Simon}},\ }\href {\doibase 10.1103/PhysRevLett.111.147202} {\bibfield
  {journal} {\bibinfo  {journal} {Phys. Rev. Lett.}\ }\textbf {\bibinfo
  {volume} {111}},\ \bibinfo {pages} {147202} (\bibinfo {year}
  {2013})}\BibitemShut {NoStop}%
\bibitem [{\citenamefont {Pientka}\ \emph {et~al.}(2013)\citenamefont
  {Pientka}, \citenamefont {Glazman},\ and\ \citenamefont {von
  Oppen}}]{Pientka2013}%
  \BibitemOpen
  \bibfield  {author} {\bibinfo {author} {\bibfnamefont {F.}~\bibnamefont
  {Pientka}}, \bibinfo {author} {\bibfnamefont {L.~I.}\ \bibnamefont
  {Glazman}}, \ and\ \bibinfo {author} {\bibfnamefont {F.}~\bibnamefont {von
  Oppen}},\ }\href {\doibase 10.1103/PhysRevB.88.155420} {\bibfield  {journal}
  {\bibinfo  {journal} {Phys. Rev. B}\ }\textbf {\bibinfo {volume} {88}},\
  \bibinfo {pages} {155420} (\bibinfo {year} {2013})}\BibitemShut {NoStop}%
\bibitem [{\citenamefont {Klinovaja}\ \emph
  {et~al.}(2013{\natexlab{a}})\citenamefont {Klinovaja}, \citenamefont {Stano},
  \citenamefont {Yazdani},\ and\ \citenamefont {Loss}}]{klinovaja.stano.13}%
  \BibitemOpen
  \bibfield  {author} {\bibinfo {author} {\bibfnamefont {J.}~\bibnamefont
  {Klinovaja}}, \bibinfo {author} {\bibfnamefont {P.}~\bibnamefont {Stano}},
  \bibinfo {author} {\bibfnamefont {A.}~\bibnamefont {Yazdani}}, \ and\
  \bibinfo {author} {\bibfnamefont {D.}~\bibnamefont {Loss}},\ }\href {\doibase
  10.1103/PhysRevLett.111.186805} {\bibfield  {journal} {\bibinfo  {journal}
  {Phys. Rev. Lett.}\ }\textbf {\bibinfo {volume} {111}},\ \bibinfo {pages}
  {186805} (\bibinfo {year} {2013}{\natexlab{a}})}\BibitemShut {NoStop}%
\bibitem [{\citenamefont {Vazifeh}\ and\ \citenamefont
  {Franz}(2013)}]{Vazifeh.Franz.13}%
  \BibitemOpen
  \bibfield  {author} {\bibinfo {author} {\bibfnamefont {M.~M.}\ \bibnamefont
  {Vazifeh}}\ and\ \bibinfo {author} {\bibfnamefont {M.}~\bibnamefont
  {Franz}},\ }\href {\doibase 10.1103/PhysRevLett.111.206802} {\bibfield
  {journal} {\bibinfo  {journal} {Phys. Rev. Lett.}\ }\textbf {\bibinfo
  {volume} {111}},\ \bibinfo {pages} {206802} (\bibinfo {year}
  {2013})}\BibitemShut {NoStop}%
\bibitem [{\citenamefont {Schecter}\ \emph {et~al.}(2016)\citenamefont
  {Schecter}, \citenamefont {Flensberg}, \citenamefont {Christensen},
  \citenamefont {Andersen},\ and\ \citenamefont {Paaske}}]{Paaske2016}%
  \BibitemOpen
  \bibfield  {author} {\bibinfo {author} {\bibfnamefont {M.}~\bibnamefont
  {Schecter}}, \bibinfo {author} {\bibfnamefont {K.}~\bibnamefont {Flensberg}},
  \bibinfo {author} {\bibfnamefont {M.}~\bibnamefont {Christensen}}, \bibinfo
  {author} {\bibfnamefont {B.}~\bibnamefont {Andersen}}, \ and\ \bibinfo
  {author} {\bibfnamefont {J.}~\bibnamefont {Paaske}},\ }\href {\doibase
  10.1103/PhysRevB.93.140503} {\bibfield  {journal} {\bibinfo  {journal} {Phys.
  Rev. B}\ }\textbf {\bibinfo {volume} {93}},\ \bibinfo {pages} {140503}
  (\bibinfo {year} {2016})}\BibitemShut {NoStop}%
\bibitem [{\citenamefont {Teixeira}\ \emph {et~al.}(2019)\citenamefont
  {Teixeira}, \citenamefont {Kuzmanovski}, \citenamefont {Black-Schaffer},\
  and\ \citenamefont {Dias~da Silva}}]{BlackSchaffer-2019}%
  \BibitemOpen
  \bibfield  {author} {\bibinfo {author} {\bibfnamefont {R.}~\bibnamefont
  {Teixeira}}, \bibinfo {author} {\bibfnamefont {D.}~\bibnamefont
  {Kuzmanovski}}, \bibinfo {author} {\bibfnamefont {A.}~\bibnamefont
  {Black-Schaffer}}, \ and\ \bibinfo {author} {\bibfnamefont {L.}~\bibnamefont
  {Dias~da Silva}},\ }\href {\doibase 10.1103/PhysRevB.99.035127} {\bibfield
  {journal} {\bibinfo  {journal} {Phys. Rev. B}\ }\textbf {\bibinfo {volume}
  {99}},\ \bibinfo {pages} {035127} (\bibinfo {year} {2019})}\BibitemShut
  {NoStop}%
\bibitem [{\citenamefont {Potter}\ and\ \citenamefont
  {Lee}(2010)}]{potter.lee.10}%
  \BibitemOpen
  \bibfield  {author} {\bibinfo {author} {\bibfnamefont {A.~C.}\ \bibnamefont
  {Potter}}\ and\ \bibinfo {author} {\bibfnamefont {P.~A.}\ \bibnamefont
  {Lee}},\ }\href {\doibase 10.1103/PhysRevLett.105.227003} {\bibfield
  {journal} {\bibinfo  {journal} {Phys. Rev. Lett.}\ }\textbf {\bibinfo
  {volume} {105}},\ \bibinfo {pages} {227003} (\bibinfo {year}
  {2010})}\BibitemShut {NoStop}%
\bibitem [{\citenamefont {Potter}\ and\ \citenamefont
  {Lee}(2011)}]{Potter2011}%
  \BibitemOpen
  \bibfield  {author} {\bibinfo {author} {\bibfnamefont {A.~C.}\ \bibnamefont
  {Potter}}\ and\ \bibinfo {author} {\bibfnamefont {P.~A.}\ \bibnamefont
  {Lee}},\ }\href {\doibase 10.1103/PhysRevB.83.094525} {\bibfield  {journal}
  {\bibinfo  {journal} {Phys. Rev. B}\ }\textbf {\bibinfo {volume} {83}},\
  \bibinfo {pages} {94525} (\bibinfo {year} {2011})}\BibitemShut {NoStop}%
\bibitem [{\citenamefont {Sedlmayr}\ \emph {et~al.}(2015)\citenamefont
  {Sedlmayr}, \citenamefont {{Aguiar-Hualde}},\ and\ \citenamefont
  {Bena}}]{Sedlmayr2015}%
  \BibitemOpen
  \bibfield  {author} {\bibinfo {author} {\bibfnamefont {N.}~\bibnamefont
  {Sedlmayr}}, \bibinfo {author} {\bibfnamefont {J.~M.}\ \bibnamefont
  {{Aguiar-Hualde}}}, \ and\ \bibinfo {author} {\bibfnamefont {C.}~\bibnamefont
  {Bena}},\ }\href {\doibase 10.1103/PhysRevB.91.115415} {\bibfield  {journal}
  {\bibinfo  {journal} {Phys. Rev. B}\ }\textbf {\bibinfo {volume} {91}},\
  \bibinfo {pages} {115415} (\bibinfo {year} {2015})}\BibitemShut {NoStop}%
\bibitem [{\citenamefont {Sedlmayr}\ \emph {et~al.}(2016)\citenamefont
  {Sedlmayr}, \citenamefont {{Aguiar-Hualde}},\ and\ \citenamefont
  {Bena}}]{Sedlmayr2016}%
  \BibitemOpen
  \bibfield  {author} {\bibinfo {author} {\bibfnamefont {N.}~\bibnamefont
  {Sedlmayr}}, \bibinfo {author} {\bibfnamefont {J.~M.}\ \bibnamefont
  {{Aguiar-Hualde}}}, \ and\ \bibinfo {author} {\bibfnamefont {C.}~\bibnamefont
  {Bena}},\ }\href {\doibase 10.1103/PhysRevB.93.155425} {\bibfield  {journal}
  {\bibinfo  {journal} {Phys. Rev. B}\ }\textbf {\bibinfo {volume} {93}},\
  \bibinfo {pages} {155425} (\bibinfo {year} {2016})}\BibitemShut {NoStop}%
\bibitem [{\citenamefont {Ptok}\ \emph {et~al.}(2017)\citenamefont {Ptok},
  \citenamefont {Kobia\l{}ka},\ and\ \citenamefont
  {Doma\'{n}ski}}]{ptok.kobialka.17}%
  \BibitemOpen
  \bibfield  {author} {\bibinfo {author} {\bibfnamefont {A.}~\bibnamefont
  {Ptok}}, \bibinfo {author} {\bibfnamefont {A.}~\bibnamefont {Kobia\l{}ka}}, \
  and\ \bibinfo {author} {\bibfnamefont {T.}~\bibnamefont {Doma\'{n}ski}},\
  }\href {\doibase 10.1103/PhysRevB.96.195430} {\bibfield  {journal} {\bibinfo
  {journal} {Phys. Rev. B}\ }\textbf {\bibinfo {volume} {96}},\ \bibinfo
  {pages} {195430} (\bibinfo {year} {2017})}\BibitemShut {NoStop}%
\bibitem [{\citenamefont {Kobia\l{}ka}\ \emph {et~al.}(2019)\citenamefont
  {Kobia\l{}ka}, \citenamefont {Doma\'{n}ski},\ and\ \citenamefont
  {Ptok}}]{kobialka.ptok.18.plaq}%
  \BibitemOpen
  \bibfield  {author} {\bibinfo {author} {\bibfnamefont {A.}~\bibnamefont
  {Kobia\l{}ka}}, \bibinfo {author} {\bibfnamefont {T.}~\bibnamefont
  {Doma\'{n}ski}}, \ and\ \bibinfo {author} {\bibfnamefont {A.}~\bibnamefont
  {Ptok}},\ }\href {\doibase 10.1038/s41598-019-49227-5} {\bibfield  {journal}
  {\bibinfo  {journal} {Sci. Rep.}\ }\textbf {\bibinfo {volume} {9}},\ \bibinfo
  {pages} {12933} (\bibinfo {year} {2019})}\BibitemShut {NoStop}%
\bibitem [{\citenamefont {Kobiałka}\ \emph {et~al.}(2019)\citenamefont
  {Kobiałka}, \citenamefont {Piekarz}, \citenamefont {Oleś},\ and\
  \citenamefont {Ptok}}]{kobialka.piekarz.20}%
  \BibitemOpen
  \bibfield  {author} {\bibinfo {author} {\bibfnamefont {A.}~\bibnamefont
  {Kobiałka}}, \bibinfo {author} {\bibfnamefont {P.}~\bibnamefont {Piekarz}},
  \bibinfo {author} {\bibfnamefont {A.~M.}\ \bibnamefont {Oleś}}, \ and\
  \bibinfo {author} {\bibfnamefont {A.}~\bibnamefont {Ptok}},\ }\href@noop {}
  {} (\bibinfo {year} {2019}),\ \Eprint {http://arxiv.org/abs/arXiv:1911.13039}
  {arXiv:1911.13039} \BibitemShut {NoStop}%
\bibitem [{\citenamefont {Kobia\l{}ka}\ and\ \citenamefont
  {Ptok}(2018)}]{kobialka.ptok.18.ring}%
  \BibitemOpen
  \bibfield  {author} {\bibinfo {author} {\bibfnamefont {A.}~\bibnamefont
  {Kobia\l{}ka}}\ and\ \bibinfo {author} {\bibfnamefont {A.}~\bibnamefont
  {Ptok}},\ }\href {\doibase 10.1088/1361-648x/ab03bf} {\bibfield  {journal}
  {\bibinfo  {journal} {J. Phys.: Condens. Matter}\ }\textbf {\bibinfo {volume}
  {31}} (\bibinfo {year} {2018}),\ 10.1088/1361-648x/ab03bf}\BibitemShut
  {NoStop}%
\bibitem [{\citenamefont {Brouwer}\ \emph {et~al.}(2011)\citenamefont
  {Brouwer}, \citenamefont {Duckheim}, \citenamefont {Romito},\ and\
  \citenamefont {von Oppen}}]{Brouwer_2011}%
  \BibitemOpen
  \bibfield  {author} {\bibinfo {author} {\bibfnamefont {P.}~\bibnamefont
  {Brouwer}}, \bibinfo {author} {\bibfnamefont {M.}~\bibnamefont {Duckheim}},
  \bibinfo {author} {\bibfnamefont {A.}~\bibnamefont {Romito}}, \ and\ \bibinfo
  {author} {\bibfnamefont {F.}~\bibnamefont {von Oppen}},\ }\href {\doibase
  10.1103/PhysRevB.84.144526} {\bibfield  {journal} {\bibinfo  {journal} {Phys.
  Rev. B}\ }\textbf {\bibinfo {volume} {84}},\ \bibinfo {pages} {144526}
  (\bibinfo {year} {2011})}\BibitemShut {NoStop}%
\bibitem [{\citenamefont {DeGottardi}\ \emph {et~al.}(2013)\citenamefont
  {DeGottardi}, \citenamefont {Sen},\ and\ \citenamefont
  {Vishveshwara}}]{DeGottardi_2013}%
  \BibitemOpen
  \bibfield  {author} {\bibinfo {author} {\bibfnamefont {W.}~\bibnamefont
  {DeGottardi}}, \bibinfo {author} {\bibfnamefont {D.}~\bibnamefont {Sen}}, \
  and\ \bibinfo {author} {\bibfnamefont {S.}~\bibnamefont {Vishveshwara}},\
  }\href {\doibase 10.1103/PhysRevLett.110.146404} {\bibfield  {journal}
  {\bibinfo  {journal} {Phys. Rev. Lett.}\ }\textbf {\bibinfo {volume} {110}},\
  \bibinfo {pages} {146404} (\bibinfo {year} {2013})}\BibitemShut {NoStop}%
\bibitem [{\citenamefont {Hui}\ \emph {et~al.}(2015)\citenamefont {Hui},
  \citenamefont {Sau},\ and\ \citenamefont {Das~Sarma}}]{Hui_2015}%
  \BibitemOpen
  \bibfield  {author} {\bibinfo {author} {\bibfnamefont {H.-Y.}\ \bibnamefont
  {Hui}}, \bibinfo {author} {\bibfnamefont {J.}~\bibnamefont {Sau}}, \ and\
  \bibinfo {author} {\bibfnamefont {S.}~\bibnamefont {Das~Sarma}},\ }\href
  {\doibase 10.1103/PhysRevB.92.174512} {\bibfield  {journal} {\bibinfo
  {journal} {Phys. Rev. B}\ }\textbf {\bibinfo {volume} {92}},\ \bibinfo
  {pages} {174512} (\bibinfo {year} {2015})}\BibitemShut {NoStop}%
\bibitem [{\citenamefont {Hoffman}\ \emph {et~al.}(2016)\citenamefont
  {Hoffman}, \citenamefont {Klinovaja},\ and\ \citenamefont
  {Loss}}]{Hoffman_2016}%
  \BibitemOpen
  \bibfield  {author} {\bibinfo {author} {\bibfnamefont {S.}~\bibnamefont
  {Hoffman}}, \bibinfo {author} {\bibfnamefont {J.}~\bibnamefont {Klinovaja}},
  \ and\ \bibinfo {author} {\bibfnamefont {D.}~\bibnamefont {Loss}},\ }\href
  {\doibase 10.1103/PhysRevB.93.165418} {\bibfield  {journal} {\bibinfo
  {journal} {Phys. Rev. B}\ }\textbf {\bibinfo {volume} {93}},\ \bibinfo
  {pages} {165418} (\bibinfo {year} {2016})}\BibitemShut {NoStop}%
\bibitem [{\citenamefont {Hegde}\ and\ \citenamefont
  {Vishveshwara}(2016)}]{Hegde_2016}%
  \BibitemOpen
  \bibfield  {author} {\bibinfo {author} {\bibfnamefont {S.~S.}\ \bibnamefont
  {Hegde}}\ and\ \bibinfo {author} {\bibfnamefont {S.}~\bibnamefont
  {Vishveshwara}},\ }\href {\doibase 10.1103/PhysRevB.94.115166} {\bibfield
  {journal} {\bibinfo  {journal} {Phys. Rev. B}\ }\textbf {\bibinfo {volume}
  {94}},\ \bibinfo {pages} {115166} (\bibinfo {year} {2016})}\BibitemShut
  {NoStop}%
\bibitem [{\citenamefont {Pekerten}\ \emph {et~al.}(2017)\citenamefont
  {Pekerten}, \citenamefont {Teker}, \citenamefont {Bozat}, \citenamefont
  {Wimmer},\ and\ \citenamefont {Adagideli}}]{Pekerten_2017}%
  \BibitemOpen
  \bibfield  {author} {\bibinfo {author} {\bibfnamefont {B.}~\bibnamefont
  {Pekerten}}, \bibinfo {author} {\bibfnamefont {A.}~\bibnamefont {Teker}},
  \bibinfo {author} {\bibfnamefont {O.}~\bibnamefont {Bozat}}, \bibinfo
  {author} {\bibfnamefont {M.}~\bibnamefont {Wimmer}}, \ and\ \bibinfo {author}
  {\bibfnamefont {i.~d.~I.}\ \bibnamefont {Adagideli}},\ }\href {\doibase
  10.1103/PhysRevB.95.064507} {\bibfield  {journal} {\bibinfo  {journal} {Phys.
  Rev. B}\ }\textbf {\bibinfo {volume} {95}},\ \bibinfo {pages} {064507}
  (\bibinfo {year} {2017})}\BibitemShut {NoStop}%
\bibitem [{\citenamefont {Ma\'{s}ka}\ \emph {et~al.}(2017)\citenamefont
  {Ma\'{s}ka}, \citenamefont {Gorczyca-Goraj}, \citenamefont {Tworzyd\l{}o},\
  and\ \citenamefont {Doma\'{n}ski}}]{maska.gorczyca.17}%
  \BibitemOpen
  \bibfield  {author} {\bibinfo {author} {\bibfnamefont {M.~M.}\ \bibnamefont
  {Ma\'{s}ka}}, \bibinfo {author} {\bibfnamefont {A.}~\bibnamefont
  {Gorczyca-Goraj}}, \bibinfo {author} {\bibfnamefont {J.}~\bibnamefont
  {Tworzyd\l{}o}}, \ and\ \bibinfo {author} {\bibfnamefont {T.}~\bibnamefont
  {Doma\'{n}ski}},\ }\href {\doibase 10.1103/PhysRevB.95.045429} {\bibfield
  {journal} {\bibinfo  {journal} {Phys. Rev. B}\ }\textbf {\bibinfo {volume}
  {95}},\ \bibinfo {pages} {045429} (\bibinfo {year} {2017})}\BibitemShut
  {NoStop}%
\bibitem [{\citenamefont {Cole}\ \emph {et~al.}(2016)\citenamefont {Cole},
  \citenamefont {Sau},\ and\ \citenamefont {Das~Sarma}}]{Cole_2016}%
  \BibitemOpen
  \bibfield  {author} {\bibinfo {author} {\bibfnamefont {W.}~\bibnamefont
  {Cole}}, \bibinfo {author} {\bibfnamefont {J.}~\bibnamefont {Sau}}, \ and\
  \bibinfo {author} {\bibfnamefont {S.}~\bibnamefont {Das~Sarma}},\ }\href
  {\doibase 10.1103/PhysRevB.94.140505} {\bibfield  {journal} {\bibinfo
  {journal} {Phys. Rev. B}\ }\textbf {\bibinfo {volume} {94}},\ \bibinfo
  {pages} {140505} (\bibinfo {year} {2016})}\BibitemShut {NoStop}%
\bibitem [{\citenamefont {Hu}\ \emph {et~al.}(2015{\natexlab{a}})\citenamefont
  {Hu}, \citenamefont {Cai}, \citenamefont {Baranov},\ and\ \citenamefont
  {Zoller}}]{Hu_2015}%
  \BibitemOpen
  \bibfield  {author} {\bibinfo {author} {\bibfnamefont {Y.}~\bibnamefont
  {Hu}}, \bibinfo {author} {\bibfnamefont {Z.}~\bibnamefont {Cai}}, \bibinfo
  {author} {\bibfnamefont {M.}~\bibnamefont {Baranov}}, \ and\ \bibinfo
  {author} {\bibfnamefont {P.}~\bibnamefont {Zoller}},\ }\href {\doibase
  10.1103/PhysRevB.92.165118} {\bibfield  {journal} {\bibinfo  {journal} {Phys.
  Rev. B}\ }\textbf {\bibinfo {volume} {92}},\ \bibinfo {pages} {165118}
  (\bibinfo {year} {2015}{\natexlab{a}})}\BibitemShut {NoStop}%
\bibitem [{\citenamefont {Klinovaja}\ and\ \citenamefont
  {Loss}(2015)}]{Klinovaja2015}%
  \BibitemOpen
  \bibfield  {author} {\bibinfo {author} {\bibfnamefont {J.}~\bibnamefont
  {Klinovaja}}\ and\ \bibinfo {author} {\bibfnamefont {D.}~\bibnamefont
  {Loss}},\ }\href {\doibase 10.1140/epjb/e2015-50882-2} {\bibfield  {journal}
  {\bibinfo  {journal} {Eur. Phys. J. B}\ }\textbf {\bibinfo {volume} {88}},\
  \bibinfo {pages} {62} (\bibinfo {year} {2015})}\BibitemShut {NoStop}%
\bibitem [{\citenamefont {Braunecker}\ and\ \citenamefont
  {Simon}(2015)}]{Simon2015}%
  \BibitemOpen
  \bibfield  {author} {\bibinfo {author} {\bibfnamefont {B.}~\bibnamefont
  {Braunecker}}\ and\ \bibinfo {author} {\bibfnamefont {P.}~\bibnamefont
  {Simon}},\ }\href {\doibase 10.1103/PhysRevB.92.241410} {\bibfield  {journal}
  {\bibinfo  {journal} {Phys. Rev. B}\ }\textbf {\bibinfo {volume} {92}},\
  \bibinfo {pages} {241410} (\bibinfo {year} {2015})}\BibitemShut {NoStop}%
\bibitem [{\citenamefont {Hu}\ \emph {et~al.}(2015{\natexlab{b}})\citenamefont
  {Hu}, \citenamefont {Scalettar},\ and\ \citenamefont
  {Singh}}]{Scalettar2015}%
  \BibitemOpen
  \bibfield  {author} {\bibinfo {author} {\bibfnamefont {W.}~\bibnamefont
  {Hu}}, \bibinfo {author} {\bibfnamefont {R.}~\bibnamefont {Scalettar}}, \
  and\ \bibinfo {author} {\bibfnamefont {R.~R.~P.}\ \bibnamefont {Singh}},\
  }\href {\doibase 10.1103/PhysRevB.92.115133} {\bibfield  {journal} {\bibinfo
  {journal} {Phys. Rev. B}\ }\textbf {\bibinfo {volume} {92}},\ \bibinfo
  {pages} {115133} (\bibinfo {year} {2015}{\natexlab{b}})}\BibitemShut
  {NoStop}%
\bibitem [{\citenamefont {Gorczyca-Goraj}\ \emph {et~al.}(2019)\citenamefont
  {Gorczyca-Goraj}, \citenamefont {Doma\'nski},\ and\ \citenamefont
  {Ma\'ska}}]{maska_etal_2019}%
  \BibitemOpen
  \bibfield  {author} {\bibinfo {author} {\bibfnamefont {A.}~\bibnamefont
  {Gorczyca-Goraj}}, \bibinfo {author} {\bibfnamefont {T.}~\bibnamefont
  {Doma\'nski}}, \ and\ \bibinfo {author} {\bibfnamefont {M.}~\bibnamefont
  {Ma\'ska}},\ }\href {\doibase 10.1103/PhysRevB.99.235430} {\bibfield
  {journal} {\bibinfo  {journal} {Phys. Rev. B}\ }\textbf {\bibinfo {volume}
  {99}},\ \bibinfo {pages} {235430} (\bibinfo {year} {2019})}\BibitemShut
  {NoStop}%
\bibitem [{\citenamefont {Kiczek}\ and\ \citenamefont
  {Ptok}(2017)}]{kiczek.ptok.17}%
  \BibitemOpen
  \bibfield  {author} {\bibinfo {author} {\bibfnamefont {B.}~\bibnamefont
  {Kiczek}}\ and\ \bibinfo {author} {\bibfnamefont {A.}~\bibnamefont {Ptok}},\
  }\href {\doibase 10.1088/1361-648X/aa93ab} {\bibfield  {journal} {\bibinfo
  {journal} {J. Phys.: Condens. Matter}\ }\textbf {\bibinfo {volume} {29}},\
  \bibinfo {pages} {495301} (\bibinfo {year} {2017})}\BibitemShut {NoStop}%
\bibitem [{\citenamefont {Kaladzhyan}\ \emph {et~al.}(2017)\citenamefont
  {Kaladzhyan}, \citenamefont {Despres}, \citenamefont {Mandal},\ and\
  \citenamefont {Bena}}]{Kaladzhyan2017a}%
  \BibitemOpen
  \bibfield  {author} {\bibinfo {author} {\bibfnamefont {V.}~\bibnamefont
  {Kaladzhyan}}, \bibinfo {author} {\bibfnamefont {J.}~\bibnamefont {Despres}},
  \bibinfo {author} {\bibfnamefont {I.}~\bibnamefont {Mandal}}, \ and\ \bibinfo
  {author} {\bibfnamefont {C.}~\bibnamefont {Bena}},\ }\href {\doibase
  10.1140/epjb/e2017-80103-y} {\bibfield  {journal} {\bibinfo  {journal} {Eur.
  Phys. J. B}\ }\textbf {\bibinfo {volume} {90}},\ \bibinfo {pages} {211}
  (\bibinfo {year} {2017})}\BibitemShut {NoStop}%
\bibitem [{\citenamefont {Wieckowski}\ \emph {et~al.}(2018)\citenamefont
  {Wieckowski}, \citenamefont {Ma\'ska},\ and\ \citenamefont
  {Mierzejewski}}]{Mierzejewski_2018}%
  \BibitemOpen
  \bibfield  {author} {\bibinfo {author} {\bibfnamefont {A.}~\bibnamefont
  {Wieckowski}}, \bibinfo {author} {\bibfnamefont {M.~M.}\ \bibnamefont
  {Ma\'ska}}, \ and\ \bibinfo {author} {\bibfnamefont {M.}~\bibnamefont
  {Mierzejewski}},\ }\href {\doibase 10.1103/PhysRevLett.120.040504} {\bibfield
   {journal} {\bibinfo  {journal} {Phys. Rev. Lett.}\ }\textbf {\bibinfo
  {volume} {120}},\ \bibinfo {pages} {040504} (\bibinfo {year}
  {2018})}\BibitemShut {NoStop}%
\bibitem [{\citenamefont {Monthus}(2018)}]{Monthus_2018}%
  \BibitemOpen
  \bibfield  {author} {\bibinfo {author} {\bibfnamefont {C.}~\bibnamefont
  {Monthus}},\ }\href {\doibase 10.1088/1751-8121/aaad14} {\bibfield  {journal}
  {\bibinfo  {journal} {J. Phys. A: Math. Theor.}\ }\textbf {\bibinfo {volume}
  {51}},\ \bibinfo {pages} {115304} (\bibinfo {year} {2018})}\BibitemShut
  {NoStop}%
\bibitem [{\citenamefont {Su}\ \emph {et~al.}(1979)\citenamefont {Su},
  \citenamefont {Schrieffer},\ and\ \citenamefont {Heeger}}]{Su_1979}%
  \BibitemOpen
  \bibfield  {author} {\bibinfo {author} {\bibfnamefont {W.~P.}\ \bibnamefont
  {Su}}, \bibinfo {author} {\bibfnamefont {J.~R.}\ \bibnamefont {Schrieffer}},
  \ and\ \bibinfo {author} {\bibfnamefont {A.~J.}\ \bibnamefont {Heeger}},\
  }\href {\doibase 10.1103/PhysRevLett.42.1698} {\bibfield  {journal} {\bibinfo
   {journal} {Phys. Rev. Lett.}\ }\textbf {\bibinfo {volume} {42}},\ \bibinfo
  {pages} {1698} (\bibinfo {year} {1979})}\BibitemShut {NoStop}%
\bibitem [{\citenamefont {Heeger}\ \emph {et~al.}(1988)\citenamefont {Heeger},
  \citenamefont {Kivelson}, \citenamefont {Schrieffer},\ and\ \citenamefont
  {Su}}]{Heeger_1988}%
  \BibitemOpen
  \bibfield  {author} {\bibinfo {author} {\bibfnamefont {A.~J.}\ \bibnamefont
  {Heeger}}, \bibinfo {author} {\bibfnamefont {S.}~\bibnamefont {Kivelson}},
  \bibinfo {author} {\bibfnamefont {J.~R.}\ \bibnamefont {Schrieffer}}, \ and\
  \bibinfo {author} {\bibfnamefont {W.~P.}\ \bibnamefont {Su}},\ }\href
  {\doibase 10.1103/RevModPhys.60.781} {\bibfield  {journal} {\bibinfo
  {journal} {Rev. Mod. Phys.}\ }\textbf {\bibinfo {volume} {60}},\ \bibinfo
  {pages} {781} (\bibinfo {year} {1988})}\BibitemShut {NoStop}%
\bibitem [{\citenamefont {Wakatsuki}\ \emph {et~al.}(2014)\citenamefont
  {Wakatsuki}, \citenamefont {Ezawa}, \citenamefont {Tanaka},\ and\
  \citenamefont {Nagaosa}}]{Wakatsuki_2014}%
  \BibitemOpen
  \bibfield  {author} {\bibinfo {author} {\bibfnamefont {R.}~\bibnamefont
  {Wakatsuki}}, \bibinfo {author} {\bibfnamefont {M.}~\bibnamefont {Ezawa}},
  \bibinfo {author} {\bibfnamefont {Y.}~\bibnamefont {Tanaka}}, \ and\ \bibinfo
  {author} {\bibfnamefont {N.}~\bibnamefont {Nagaosa}},\ }\href {\doibase
  10.1103/PhysRevB.90.014505} {\bibfield  {journal} {\bibinfo  {journal} {Phys.
  Rev. B}\ }\textbf {\bibinfo {volume} {90}},\ \bibinfo {pages} {014505}
  (\bibinfo {year} {2014})}\BibitemShut {NoStop}%
\bibitem [{\citenamefont {Wang}\ \emph {et~al.}(2017)\citenamefont {Wang},
  \citenamefont {Miao}, \citenamefont {Jin},\ and\ \citenamefont
  {Chen}}]{Wang_2017}%
  \BibitemOpen
  \bibfield  {author} {\bibinfo {author} {\bibfnamefont {Y.}~\bibnamefont
  {Wang}}, \bibinfo {author} {\bibfnamefont {J.-J.}\ \bibnamefont {Miao}},
  \bibinfo {author} {\bibfnamefont {H.-K.}\ \bibnamefont {Jin}}, \ and\
  \bibinfo {author} {\bibfnamefont {S.}~\bibnamefont {Chen}},\ }\href {\doibase
  10.1103/PhysRevB.96.205428} {\bibfield  {journal} {\bibinfo  {journal} {Phys.
  Rev. B}\ }\textbf {\bibinfo {volume} {96}},\ \bibinfo {pages} {205428}
  (\bibinfo {year} {2017})}\BibitemShut {NoStop}%
\bibitem [{\citenamefont {Ezawa}(2017)}]{Ezawa_2017}%
  \BibitemOpen
  \bibfield  {author} {\bibinfo {author} {\bibfnamefont {M.}~\bibnamefont
  {Ezawa}},\ }\href {\doibase 10.1103/PhysRevB.96.121105} {\bibfield  {journal}
  {\bibinfo  {journal} {Phys. Rev. B}\ }\textbf {\bibinfo {volume} {96}},\
  \bibinfo {pages} {121105} (\bibinfo {year} {2017})}\BibitemShut {NoStop}%
\bibitem [{\citenamefont {Chitov}(2018)}]{Chitov_2018}%
  \BibitemOpen
  \bibfield  {author} {\bibinfo {author} {\bibfnamefont {G.}~\bibnamefont
  {Chitov}},\ }\href {\doibase 10.1103/PhysRevB.97.085131} {\bibfield
  {journal} {\bibinfo  {journal} {Phys. Rev. B}\ }\textbf {\bibinfo {volume}
  {97}},\ \bibinfo {pages} {085131} (\bibinfo {year} {2018})}\BibitemShut
  {NoStop}%
\bibitem [{\citenamefont {Yu}\ \emph {et~al.}(2019)\citenamefont {Yu},
  \citenamefont {Sacramento}, \citenamefont {Li}, \citenamefont {Angelakis},\
  and\ \citenamefont {Lin}}]{Yu_2019}%
  \BibitemOpen
  \bibfield  {author} {\bibinfo {author} {\bibfnamefont {W.}~\bibnamefont
  {Yu}}, \bibinfo {author} {\bibfnamefont {P.}~\bibnamefont {Sacramento}},
  \bibinfo {author} {\bibfnamefont {Y.}~\bibnamefont {Li}}, \bibinfo {author}
  {\bibfnamefont {D.}~\bibnamefont {Angelakis}}, \ and\ \bibinfo {author}
  {\bibfnamefont {H.-Q.}\ \bibnamefont {Lin}},\ }\href {\doibase
  10.1103/PhysRevB.99.115113} {\bibfield  {journal} {\bibinfo  {journal} {Phys.
  Rev. B}\ }\textbf {\bibinfo {volume} {99}},\ \bibinfo {pages} {115113}
  (\bibinfo {year} {2019})}\BibitemShut {NoStop}%
\bibitem [{\citenamefont {Hua}\ \emph {et~al.}(2019)\citenamefont {Hua},
  \citenamefont {Chen}, \citenamefont {Xu},\ and\ \citenamefont
  {Zhou}}]{Hua_2019}%
  \BibitemOpen
  \bibfield  {author} {\bibinfo {author} {\bibfnamefont {C.-B.}\ \bibnamefont
  {Hua}}, \bibinfo {author} {\bibfnamefont {R.}~\bibnamefont {Chen}}, \bibinfo
  {author} {\bibfnamefont {D.-H.}\ \bibnamefont {Xu}}, \ and\ \bibinfo {author}
  {\bibfnamefont {B.}~\bibnamefont {Zhou}},\ }\href {\doibase
  10.1103/PhysRevB.100.205302} {\bibfield  {journal} {\bibinfo  {journal}
  {Phys. Rev. B}\ }\textbf {\bibinfo {volume} {100}},\ \bibinfo {pages}
  {205302} (\bibinfo {year} {2019})}\BibitemShut {NoStop}%
\bibitem [{\citenamefont {Moore}\ and\ \citenamefont
  {Balents}(2007)}]{moore.balents.07}%
  \BibitemOpen
  \bibfield  {author} {\bibinfo {author} {\bibfnamefont {J.~E.}\ \bibnamefont
  {Moore}}\ and\ \bibinfo {author} {\bibfnamefont {L.}~\bibnamefont
  {Balents}},\ }\href {\doibase 10.1103/PhysRevB.75.121306} {\bibfield
  {journal} {\bibinfo  {journal} {Phys. Rev. B}\ }\textbf {\bibinfo {volume}
  {75}},\ \bibinfo {pages} {121306} (\bibinfo {year} {2007})}\BibitemShut
  {NoStop}%
\bibitem [{\citenamefont {Menzel}\ \emph {et~al.}(2012)\citenamefont {Menzel},
  \citenamefont {Mokrousov}, \citenamefont {Wieser}, \citenamefont {Bickel},
  \citenamefont {Vedmedenko}, \citenamefont {Bl\"ugel}, \citenamefont {Heinze},
  \citenamefont {von Bergmann}, \citenamefont {Kubetzka},\ and\ \citenamefont
  {Wiesendanger}}]{Wiesendanger-2012}%
  \BibitemOpen
  \bibfield  {author} {\bibinfo {author} {\bibfnamefont {M.}~\bibnamefont
  {Menzel}}, \bibinfo {author} {\bibfnamefont {Y.}~\bibnamefont {Mokrousov}},
  \bibinfo {author} {\bibfnamefont {R.}~\bibnamefont {Wieser}}, \bibinfo
  {author} {\bibfnamefont {J.~E.}\ \bibnamefont {Bickel}}, \bibinfo {author}
  {\bibfnamefont {E.}~\bibnamefont {Vedmedenko}}, \bibinfo {author}
  {\bibfnamefont {S.}~\bibnamefont {Bl\"ugel}}, \bibinfo {author}
  {\bibfnamefont {S.}~\bibnamefont {Heinze}}, \bibinfo {author} {\bibfnamefont
  {K.}~\bibnamefont {von Bergmann}}, \bibinfo {author} {\bibfnamefont
  {A.}~\bibnamefont {Kubetzka}}, \ and\ \bibinfo {author} {\bibfnamefont
  {R.}~\bibnamefont {Wiesendanger}},\ }\href {\doibase
  10.1103/PhysRevLett.108.197204} {\bibfield  {journal} {\bibinfo  {journal}
  {Phys. Rev. Lett.}\ }\textbf {\bibinfo {volume} {108}},\ \bibinfo {pages}
  {197204} (\bibinfo {year} {2012})}\BibitemShut {NoStop}%
\bibitem [{\citenamefont {Nadj-Perge}\ \emph
  {et~al.}(2013{\natexlab{b}})\citenamefont {Nadj-Perge}, \citenamefont
  {Drozdov}, \citenamefont {Bernevig},\ and\ \citenamefont
  {Yazdani}}]{Yazdani-2013}%
  \BibitemOpen
  \bibfield  {author} {\bibinfo {author} {\bibfnamefont {S.}~\bibnamefont
  {Nadj-Perge}}, \bibinfo {author} {\bibfnamefont {I.~K.}\ \bibnamefont
  {Drozdov}}, \bibinfo {author} {\bibfnamefont {B.~A.}\ \bibnamefont
  {Bernevig}}, \ and\ \bibinfo {author} {\bibfnamefont {A.}~\bibnamefont
  {Yazdani}},\ }\href {\doibase 10.1103/PhysRevB.88.020407} {\bibfield
  {journal} {\bibinfo  {journal} {Phys. Rev. B}\ }\textbf {\bibinfo {volume}
  {88}},\ \bibinfo {pages} {020407} (\bibinfo {year}
  {2013}{\natexlab{b}})}\BibitemShut {NoStop}%
\bibitem [{\citenamefont {Klinovaja}\ \emph
  {et~al.}(2013{\natexlab{b}})\citenamefont {Klinovaja}, \citenamefont {Stano},
  \citenamefont {Yazdani},\ and\ \citenamefont {Loss}}]{Klinovaja-2013}%
  \BibitemOpen
  \bibfield  {author} {\bibinfo {author} {\bibfnamefont {J.}~\bibnamefont
  {Klinovaja}}, \bibinfo {author} {\bibfnamefont {P.}~\bibnamefont {Stano}},
  \bibinfo {author} {\bibfnamefont {A.}~\bibnamefont {Yazdani}}, \ and\
  \bibinfo {author} {\bibfnamefont {D.}~\bibnamefont {Loss}},\ }\href {\doibase
  10.1103/PhysRevLett.111.186805} {\bibfield  {journal} {\bibinfo  {journal}
  {Phys. Rev. Lett.}\ }\textbf {\bibinfo {volume} {111}},\ \bibinfo {pages}
  {186805} (\bibinfo {year} {2013}{\natexlab{b}})}\BibitemShut {NoStop}%
\bibitem [{\citenamefont {Wiesendanger}(2009)}]{wiesendanger.09}%
  \BibitemOpen
  \bibfield  {author} {\bibinfo {author} {\bibfnamefont {R.}~\bibnamefont
  {Wiesendanger}},\ }\href {\doibase 10.1103/RevModPhys.81.1495} {\bibfield
  {journal} {\bibinfo  {journal} {Rev. Mod. Phys.}\ }\textbf {\bibinfo {volume}
  {81}},\ \bibinfo {pages} {1495} (\bibinfo {year} {2009})}\BibitemShut
  {NoStop}%
\bibitem [{\citenamefont {Figgins}\ and\ \citenamefont
  {Morr}(2010)}]{figgins.morr.10}%
  \BibitemOpen
  \bibfield  {author} {\bibinfo {author} {\bibfnamefont {J.}~\bibnamefont
  {Figgins}}\ and\ \bibinfo {author} {\bibfnamefont {D.~K.}\ \bibnamefont
  {Morr}},\ }\href {\doibase 10.1103/PhysRevLett.104.187202} {\bibfield
  {journal} {\bibinfo  {journal} {Phys. Rev. Lett.}\ }\textbf {\bibinfo
  {volume} {104}},\ \bibinfo {pages} {187202} (\bibinfo {year}
  {2010})}\BibitemShut {NoStop}%
\bibitem [{\citenamefont {Ryu}\ \emph {et~al.}(2010)\citenamefont {Ryu},
  \citenamefont {Schnyder}, \citenamefont {Furusaki},\ and\ \citenamefont
  {Ludwig}}]{Ryu2010}%
  \BibitemOpen
  \bibfield  {author} {\bibinfo {author} {\bibfnamefont {S.}~\bibnamefont
  {Ryu}}, \bibinfo {author} {\bibfnamefont {A.}~\bibnamefont {Schnyder}},
  \bibinfo {author} {\bibfnamefont {A.}~\bibnamefont {Furusaki}}, \ and\
  \bibinfo {author} {\bibfnamefont {A.}~\bibnamefont {Ludwig}},\ }\href
  {\doibase 10.1088/1367-2630/12/6/065010} {\bibfield  {journal} {\bibinfo
  {journal} {New J. Phys.}\ }\textbf {\bibinfo {volume} {12}},\ \bibinfo
  {pages} {65010} (\bibinfo {year} {2010})}\BibitemShut {NoStop}%
\bibitem [{\citenamefont {Akhmerov}\ \emph {et~al.}(2011)\citenamefont
  {Akhmerov}, \citenamefont {Dahlhaus}, \citenamefont {Hassler}, \citenamefont
  {Wimmer},\ and\ \citenamefont {Beenakker}}]{akhmerov.11}%
  \BibitemOpen
  \bibfield  {author} {\bibinfo {author} {\bibfnamefont {A.~R.}\ \bibnamefont
  {Akhmerov}}, \bibinfo {author} {\bibfnamefont {J.~P.}\ \bibnamefont
  {Dahlhaus}}, \bibinfo {author} {\bibfnamefont {F.}~\bibnamefont {Hassler}},
  \bibinfo {author} {\bibfnamefont {M.}~\bibnamefont {Wimmer}}, \ and\ \bibinfo
  {author} {\bibfnamefont {C.~W.~J.}\ \bibnamefont {Beenakker}},\ }\href
  {\doibase 10.1103/PhysRevLett.106.057001} {\bibfield  {journal} {\bibinfo
  {journal} {Phys. Rev. Lett.}\ }\textbf {\bibinfo {volume} {106}},\ \bibinfo
  {pages} {057001} (\bibinfo {year} {2011})}\BibitemShut {NoStop}%
\bibitem [{\citenamefont {Fulga}\ \emph {et~al.}(2011)\citenamefont {Fulga},
  \citenamefont {Hassler}, \citenamefont {Akhmerov},\ and\ \citenamefont
  {Beenakker}}]{fulga.11}%
  \BibitemOpen
  \bibfield  {author} {\bibinfo {author} {\bibfnamefont {I.~C.}\ \bibnamefont
  {Fulga}}, \bibinfo {author} {\bibfnamefont {F.}~\bibnamefont {Hassler}},
  \bibinfo {author} {\bibfnamefont {A.~R.}\ \bibnamefont {Akhmerov}}, \ and\
  \bibinfo {author} {\bibfnamefont {C.~W.~J.}\ \bibnamefont {Beenakker}},\
  }\href {\doibase 10.1103/PhysRevB.83.155429} {\bibfield  {journal} {\bibinfo
  {journal} {Phys. Rev. B}\ }\textbf {\bibinfo {volume} {83}},\ \bibinfo
  {pages} {155429} (\bibinfo {year} {2011})}\BibitemShut {NoStop}%
\bibitem [{\citenamefont {Sato}(2009)}]{Sato2009b}%
  \BibitemOpen
  \bibfield  {author} {\bibinfo {author} {\bibfnamefont {M.}~\bibnamefont
  {Sato}},\ }\href {\doibase 10.1103/PhysRevB.79.214526} {\bibfield  {journal}
  {\bibinfo  {journal} {Phys. Rev. B}\ }\textbf {\bibinfo {volume} {79}},\
  \bibinfo {pages} {214526} (\bibinfo {year} {2009})}\BibitemShut {NoStop}%
\bibitem [{\citenamefont {Dutreix}(2017)}]{Dutreix2017}%
  \BibitemOpen
  \bibfield  {author} {\bibinfo {author} {\bibfnamefont {C.}~\bibnamefont
  {Dutreix}},\ }\href {\doibase 10.1103/PhysRevB.96.045416} {\bibfield
  {journal} {\bibinfo  {journal} {Phys. Rev. B}\ }\textbf {\bibinfo {volume}
  {96}},\ \bibinfo {pages} {45416} (\bibinfo {year} {2017})}\BibitemShut
  {NoStop}%
\bibitem [{\citenamefont {Sedlmayr}\ and\ \citenamefont
  {Bena}(2015)}]{Sedlmayr2015b}%
  \BibitemOpen
  \bibfield  {author} {\bibinfo {author} {\bibfnamefont {N.}~\bibnamefont
  {Sedlmayr}}\ and\ \bibinfo {author} {\bibfnamefont {C.}~\bibnamefont
  {Bena}},\ }\href {\doibase 10.1103/PhysRevB.92.115115} {\bibfield  {journal}
  {\bibinfo  {journal} {Phys. Rev. B}\ }\textbf {\bibinfo {volume} {92}},\
  \bibinfo {pages} {115115} (\bibinfo {year} {2015})}\BibitemShut {NoStop}%
\bibitem [{\citenamefont {Sedlmayr}\ \emph {et~al.}(2017)\citenamefont
  {Sedlmayr}, \citenamefont {Kaladzhyan}, \citenamefont {Dutreix},\ and\
  \citenamefont {Bena}}]{Sedlmayr2017}%
  \BibitemOpen
  \bibfield  {author} {\bibinfo {author} {\bibfnamefont {N.}~\bibnamefont
  {Sedlmayr}}, \bibinfo {author} {\bibfnamefont {V.}~\bibnamefont
  {Kaladzhyan}}, \bibinfo {author} {\bibfnamefont {C.}~\bibnamefont {Dutreix}},
  \ and\ \bibinfo {author} {\bibfnamefont {C.}~\bibnamefont {Bena}},\ }\href
  {\doibase 10.1103/PhysRevB.96.184516} {\bibfield  {journal} {\bibinfo
  {journal} {Phys. Rev. B}\ }\textbf {\bibinfo {volume} {96}},\ \bibinfo
  {pages} {184516} (\bibinfo {year} {2017})}\BibitemShut {NoStop}%
\bibitem [{\citenamefont {Franca}\ \emph {et~al.}(2019)\citenamefont {Franca},
  \citenamefont {Efremov},\ and\ \citenamefont {Fulga}}]{Fulga-2019}%
  \BibitemOpen
  \bibfield  {author} {\bibinfo {author} {\bibfnamefont {S.}~\bibnamefont
  {Franca}}, \bibinfo {author} {\bibfnamefont {D.~V.}\ \bibnamefont {Efremov}},
  \ and\ \bibinfo {author} {\bibfnamefont {I.~C.}\ \bibnamefont {Fulga}},\
  }\href {\doibase 10.1103/PhysRevB.100.075415} {\bibfield  {journal} {\bibinfo
   {journal} {Phys. Rev. B}\ }\textbf {\bibinfo {volume} {100}},\ \bibinfo
  {pages} {075415} (\bibinfo {year} {2019})}\BibitemShut {NoStop}%
\bibitem [{\citenamefont {Drost}\ \emph {et~al.}(2017)\citenamefont {Drost},
  \citenamefont {Ojanen}, \citenamefont {Harju},\ and\ \citenamefont
  {Liljeroth}}]{Drost-2017}%
  \BibitemOpen
  \bibfield  {author} {\bibinfo {author} {\bibfnamefont {R.}~\bibnamefont
  {Drost}}, \bibinfo {author} {\bibfnamefont {T.}~\bibnamefont {Ojanen}},
  \bibinfo {author} {\bibfnamefont {A.}~\bibnamefont {Harju}}, \ and\ \bibinfo
  {author} {\bibfnamefont {P.}~\bibnamefont {Liljeroth}},\ }\href {\doibase
  10.1038/nphys4080} {\bibfield  {journal} {\bibinfo  {journal} {Nat. Phys.}\
  }\textbf {\bibinfo {volume} {13}},\ \bibinfo {pages} {668} (\bibinfo {year}
  {2017})}\BibitemShut {NoStop}%
\bibitem [{\citenamefont {Gonz\'alez-Cuadra}\ \emph {et~al.}(2018)\citenamefont
  {Gonz\'alez-Cuadra}, \citenamefont {Grzybowski}, \citenamefont {Dauphin},\
  and\ \citenamefont {Lewenstein}}]{Lewenstein-PRL2019}%
  \BibitemOpen
  \bibfield  {author} {\bibinfo {author} {\bibfnamefont {D.}~\bibnamefont
  {Gonz\'alez-Cuadra}}, \bibinfo {author} {\bibfnamefont {P.}~\bibnamefont
  {Grzybowski}}, \bibinfo {author} {\bibfnamefont {A.}~\bibnamefont {Dauphin}},
  \ and\ \bibinfo {author} {\bibfnamefont {M.}~\bibnamefont {Lewenstein}},\
  }\href {\doibase 10.1103/PhysRevLett.121.090402} {\bibfield  {journal}
  {\bibinfo  {journal} {Phys. Rev. Lett.}\ }\textbf {\bibinfo {volume} {121}},\
  \bibinfo {pages} {090402} (\bibinfo {year} {2018})}\BibitemShut {NoStop}%
\bibitem [{\citenamefont {Gonz\'alez-Cuadra}\ \emph {et~al.}(2019)\citenamefont
  {Gonz\'alez-Cuadra}, \citenamefont {Dauphin}, \citenamefont {Grzybowski},
  \citenamefont {W\'ojcik}, \citenamefont {Lewenstein},\ and\ \citenamefont
  {Bermudez}}]{Lewenstein-PRB2019}%
  \BibitemOpen
  \bibfield  {author} {\bibinfo {author} {\bibfnamefont {D.}~\bibnamefont
  {Gonz\'alez-Cuadra}}, \bibinfo {author} {\bibfnamefont {A.}~\bibnamefont
  {Dauphin}}, \bibinfo {author} {\bibfnamefont {P.}~\bibnamefont {Grzybowski}},
  \bibinfo {author} {\bibfnamefont {P.}~\bibnamefont {W\'ojcik}}, \bibinfo
  {author} {\bibfnamefont {M.}~\bibnamefont {Lewenstein}}, \ and\ \bibinfo
  {author} {\bibfnamefont {A.}~\bibnamefont {Bermudez}},\ }\href {\doibase
  10.1103/PhysRevB.99.045139} {\bibfield  {journal} {\bibinfo  {journal} {Phys.
  Rev. B}\ }\textbf {\bibinfo {volume} {99}},\ \bibinfo {pages} {045139}
  (\bibinfo {year} {2019})}\BibitemShut {NoStop}%
\end{thebibliography}%

\end{document}